\documentclass[prl,a4paper,showpacs,preprintnumbers,amsmath,amssymb%
nobibnotes,nofootinbib,flloatfix,amstex]{revtex4}
\newcommand{\ket}[1]{|\,#1\,\rangle}                %
\newcommand{\bra}[1]{\langle\,#1\,}                 %
\newfont{\Bb}{msbm10}                   %
\usepackage{graphicx,graphics,wrapfig,rotating} 
\usepackage{dcolumn}                
\usepackage{bm}                 

\begin{document}

\title{COARSE GRAINED LIOUVILLE DYNAMICS OF PIECEWISE LINEAR DISCONTINUOUS MAPS}
\author{M.E.Spina $^{a}$}
\email{spina@tandar.cnea.gov.ar}

\author{M. Saraceno $^{a,b}$ }
\email{saraceno@tandar.cnea.gov.ar}

 \affiliation{%
$^{a}$ Dto. de F\'\i sica, Comisi\'on Nacional de Energ\'\i a At\'omica.
Libertador 8250 (1429), Buenos Aires, Argentina.\\
$^{b}$Escuela de Ciencia y Tecnolog\'\i a, Universidad Nacional de San
Mart\'\i n. Alem 3901 (B1653HIM), Villa Ballester, Provincia de  Buenos
Aires,
Argentina.}%

\date{\today}

\begin{abstract}
\noindent We compute the spectrum of the classical and quantum
mechanical coarse-grained propagators for a piecewise linear
discontinuous map.
We analyze the quantum - classical correspondence and
the evolution of the spectrum with increasing resolution.
Our results are compared to the ones obtained for a
mixed system.
\end{abstract}

\maketitle

\begin{figure}[h]
\begin{center}
\includegraphics*[width=8cm,angle=-90]{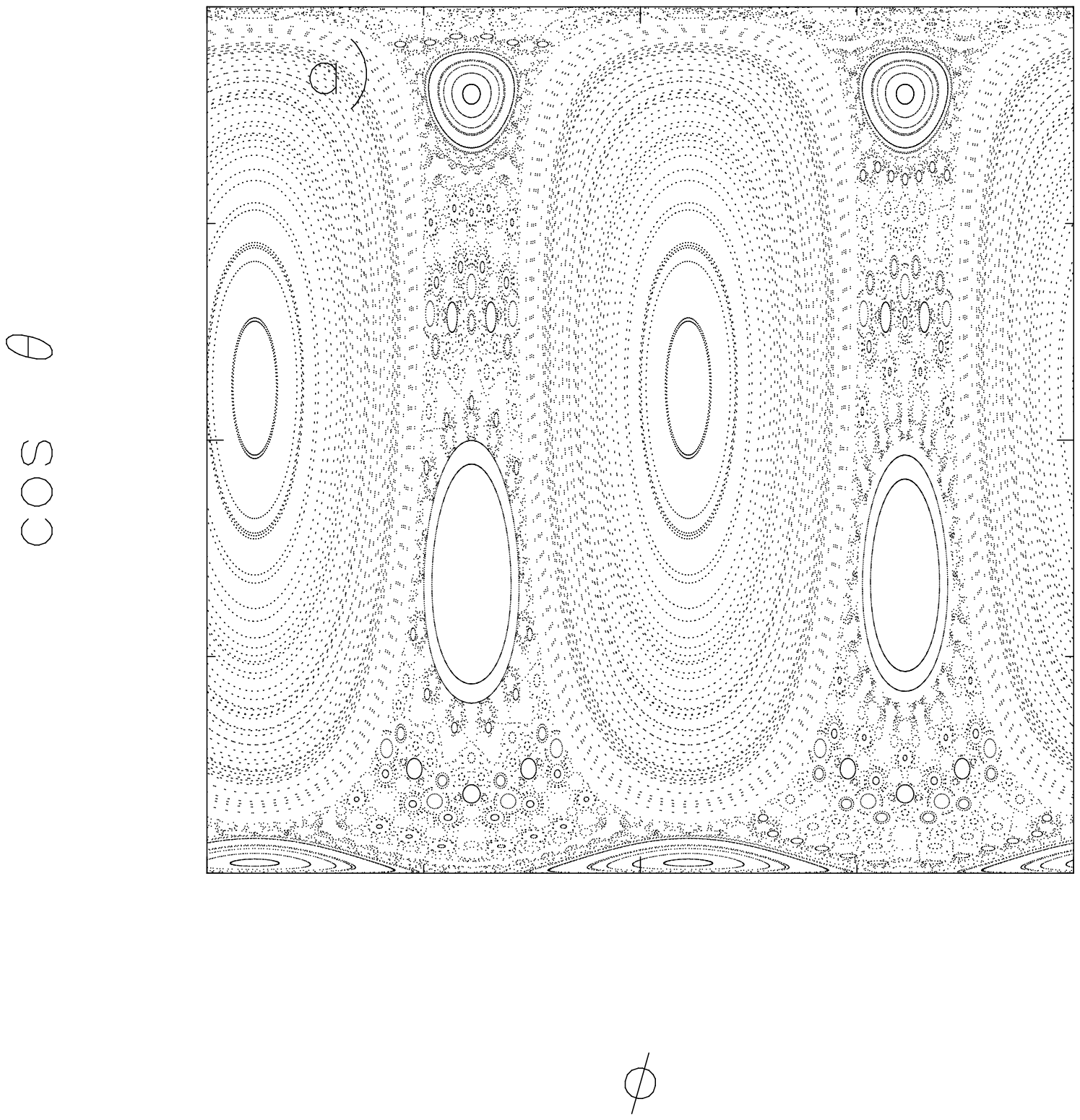}
\vspace{0.2cm}
\includegraphics*[width=8cm,angle=-90]{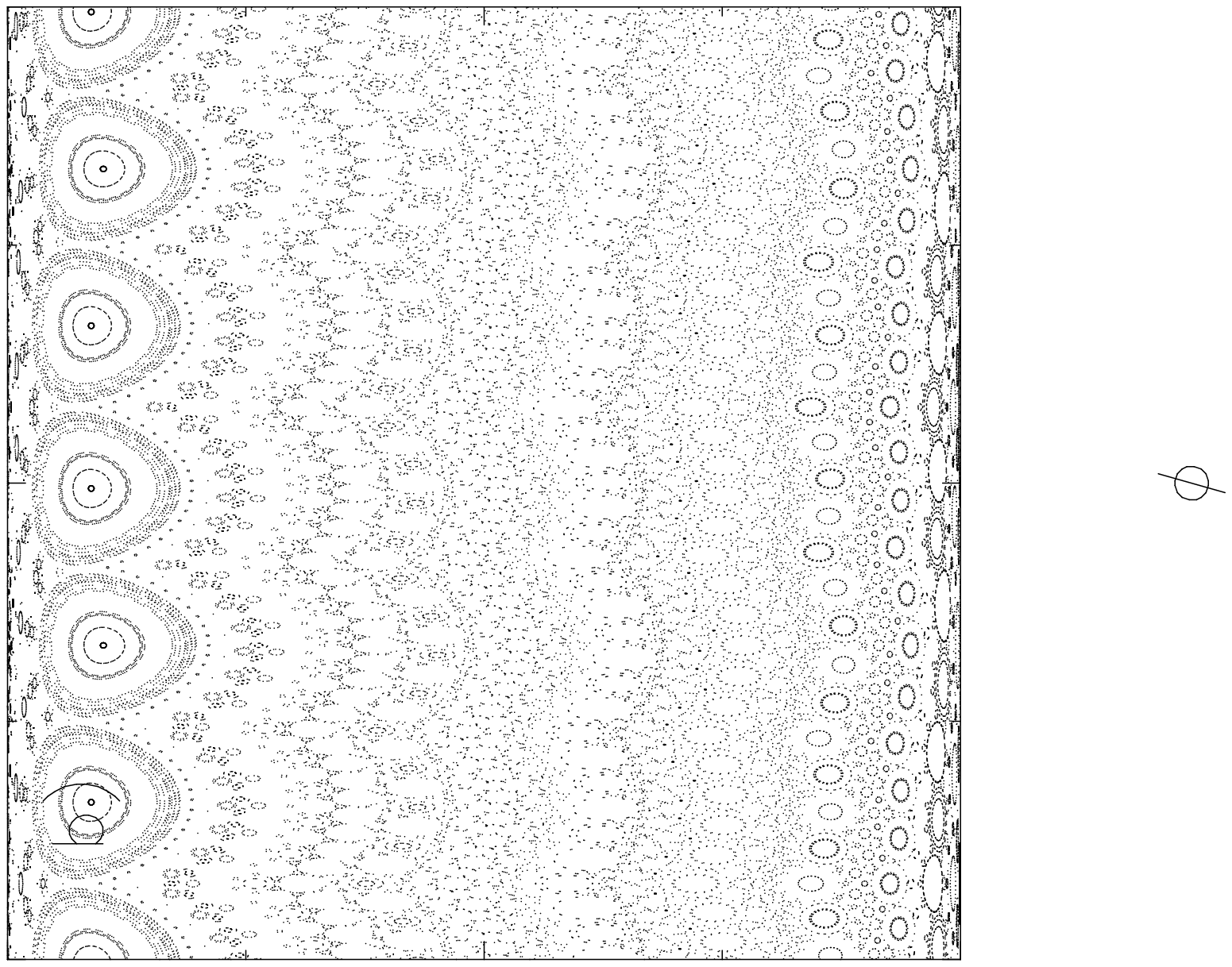}
\vspace{0.2cm}
\caption{%
Phase space portrait of map eq.(\ref{map1}) for a) $ \omega=  \mu=
\pi (\sqrt{5}-1)$) ( Case I), b) $ \omega= 1 \, \mu= 0.05 $( Case
II). \label{spect0} }
\end{center}
\end{figure}

In this work we consider systems with a highly complex regular
dynamics exhibiting, as in the chaotic case, an endless hierarchy
of phase space structures. Examples of such dynamics are found in
polygonal billiards on a surface with spherical curvature
\cite{Spi1} and also in a class of maps introduced in
\cite{Sco1,Sco2} that can be characterized as discontinuous
piecewise linear. Typical phase space portraits of such maps are
shown in Fig. (\ref{spect0}). An infinite number of chains of
elliptic islands pack the phase space. In the center of each
island sits a stable periodic orbit of arbitrarily long period
surrounded by a family of nested invariant curves which correspond
to open orbits having an infinite but periodic sequence with the
same period. The motion inside each island is regular. Their size
decreases as the period increases and phase space takes a fractal
structure. The boundary of the islands constitute the unstable set
which is the infinite iteration of the discontinuity. The motion
in these systems is non chaotic but highly irregular. In the case
of the billiards this peculiar phase space is a consequence of the
focusing mechanism on the sphere, while in the second case it is
due to the discontinuity of the map.

The corresponding quantum mechanical systems have been
investigated in both examples. For the billiards it was shown
\cite{Spi2} that the quantum energy spectra follow a non universal
intermediate statistics that can be understood in reference to the
classical phase space plot, but no quantitative classical-quantum
mechanical correspondence was derived. In \cite{Sco2} the
semiclassical regime of a piecewise linear map was explored using
Gutzwiller's method of periodic orbit quantization. However, as
pointed out by the authors, great difficulties arose in their
semiclassical approximation when describing the eigenstates not
supported by stable periodic orbits.  The relative failure of
Gutzwiller's approach was attributed to the ellipticity of the
dynamics of the map.

In the present contribution we attempt a different approach. We
consider the dicontinuous mapping introduced in \cite{Sco2} from
the point of view of the spectral properties of the Liouville
dynamics and study both the classical and the quantum Liouvillian
at limited phase space resolution. The motivation is to compare
the results to those well known that apply to chaotic or mixed
systems \cite{Haa1,Haa2}.

In pure hyperbolic systems the asymptotic decay of classical
correlation functions is exponential and the decay rates can be
rigorously obtained from the Ruelle-Pollicott ( RP ) resonances
\cite{Ru}.
To study the long time behavior of general chaotic and mixed
systems, which are beyond the validity of the RP theorem, non
rigorous methods have been developed to compute the resonances of
the Frobenius-Perron(FP)propagator.
 All these approaches are based
 on a coarse grained Liouville dynamics of the density function in the limit of zero coarse
 graining \cite{Fi}.
 In \cite{Na} the blurring of phase space structures is implemented by adding a diffusive noise
 in the Liouville equation, which results in a coarse graining of the FP propagator and
  the limit of vanishing noise is finally considered .
 An alternative method developed in \cite{Haa1,Haa2} uses a truncation of the infinite unitary FP operator
  to a finite dimension $ N $ in a basis of functions ordered by increasing resolution. The eigenvalues of this non
  unitary $ N X N $ operator are calculated and
   the the limit $ N \rightarrow \infty $ is taken.
Since noisy propagation restores the quantum-classical
correspondence, which is otherwise
 lost in short time scales, the computation of RP resonances for chaotic systems is a powerful tool to explore
 the link between classical and quantum mechanical dynamics and to look for the emergence of chaotic signatures in the
  quantum mechanical system.

In the present work we compute the spectrum
of the classical coarse grained FP and quantum ( Husimi) propagator for the discontinuous map on the sphere
presented in \cite{Sco2}, using the truncation method of \cite{Haa1,Haa2}. We follow
closely the procedure that Haake and collaborators developed to study a system with mixed dynamics: the kicked top.
We show that for low resolution classical and quantum mechanical propagators coincide, although the convergence is slow.
We study the behavior of the eigenvalues and eigenfunctions of the truncated propagator with increasing resolution
and compare our results to the ones obtained for a map with a mixed phase space \cite{Haa1}.

The area preserving map considered in  \cite{Sco2} acts on the angular momentum vector
$ \hat{J} = (J_x,J_y,J_z) =j \ (\sin \theta  \ \cos \phi, \sin \theta \ \sin \phi ,\cos \theta) $ of fixed length $ j $.
The corresponding phase space is the sphere, and $ \cos \theta $ and $ \phi $ the canonical variables.
It has the form:
\begin{equation}
M = R_z(\omega) \ |R_x(\mu)| \label{map1}
\end{equation}

\noindent that can also be rewritten as:

\begin{equation}
M = R_z(\omega) \ R_y(-\pi /2) \ |R_z(\mu)| \ R_y(\pi /2)
\label{map2}
\end{equation}

\noindent where $ |R_i(\mu)| $ denotes a rotation around the $ J_i $ axis, of angle $ \mu $  sign $ J_i $.
Under the action of this map the points on the sphere with $ J_x < 0 $  rotate in $ \mu $ while those with
$ J_x > 0 $ in $ - \mu $ around the $ J_x $ axis.  The whole sphere rotates then in $ \omega $ around the $ J_z $
axis. That is, at each step, every point rotates linearly except for the ones on the great circle $ J_x = 0 $
 where the map becomes singular. The unstable set is constituted by all the images and preimages
of this great circle.
This is evidenced in the phase space portrait. The bounds of one island of each chain are tangent to the
 unstability line $ \phi= \pi/2, 3 \pi /2 $.

The evolution of the phase space density $ \rho $ of the system is given by:

\begin{equation}
\rho_{n+1}(\cos \theta,\phi) = { \cal P } \ \rho_{n}(\cos
\theta,\phi) \label{fp}
\end{equation}

\noindent where $ { \cal P } $ is the FP propagator that is related to the invertible map (\ref{map1}) by:

\begin{equation}
{ \cal P } \ \rho(\cos \theta,\phi)=  \rho(M^{-1}(\cos
\theta,\phi))
\end{equation}

Eq.(\ref{fp}) results from the formal integration of the Liouville equation:
\begin{equation}
{\partial}_t \ \rho = {\cal L} \ \rho
\end{equation}

\noindent Defining the Liouvillians corresponding to each rotation
in eq.(\ref{map2}), the FP operator corresponding to this map
reads $ { \cal P } = \exp[ {\cal L}_{R_z(\omega)}] \ \exp[ {\cal
L}_{R_y(-\pi /2)}] \ \exp[ {\cal L}_{|R_z(\mu)|}] \
 \exp[ {\cal L}_{R_y(\pi /2)}] $ .

\noindent The matrix elements of $ { \cal P }$ can be computed in the basis of the spherical
harmonics $Y_{l m} (\theta, \phi)$, which span
the space of phase space functions on the sphere and are ordered according to increasing resolution by index $ l $.

It can be easily shown that:

\begin{equation}
(\exp[ {\cal L}_{R_z(\omega)}])_{l m,l' m'}=\delta_{l,l'}
\delta_{m,m'} \ \exp[-i m \omega] \label{class1}
\end{equation}

\begin{equation}
(\exp[ {\cal L}_{R_y(\omega)}])_{l m,l',m'}=\delta_{l,l'} \
d^{l}_{m,m'}(\omega) \label{class2}
\end{equation}

\begin{eqnarray}
&& (\exp[ {\cal L}_|{R_z(\mu)|}])_{l m,l',m'}= \nonumber\\
&&  \int_{0}^{2 \pi} d \phi \
 \{ \int_{-1}^0 d\cos \theta \
Y^*_{l m}(\theta, \phi) \ Y_{l' m'}(\theta, \phi-\mu) + \int_{0}^1
d\cos \theta \ Y^*_{l m}(\theta, \phi) \ Y_{l' m'}(\theta,
\phi+\mu) \} \nonumber
\end{eqnarray}

\begin{eqnarray}
= \delta_{m,m'}
\left\{
\begin{array} {ccc}
   \cos \mu & \mbox{for} \quad   l=l' \\
   0 & \mbox{for even} \quad l-l'\\
   i/2 \ \sin m \mu \int_{-1}^0 d\cos \theta \ P_{l,m}(\cos \theta) \ P_{l',m'}(\cos \theta) & \mbox{for odd} \quad l-l'
\end{array}
\right.
\label{class3}
\end{eqnarray}

\noindent where $ d^{l}_{m,m'}(\omega)$ are the Wigner's d-matrix and $ P_{l,m}(\cos \theta)$
the Legendre functions.
The last matrix element, which can be computed analytically, couples subspaces with different values
of $ l $. Therefore, a truncation of $ \cal P $ up to a finite $ l_{max} $ leads to a non unitary matrix.

 The quantum version of the map in the Hilbert space of the wave functions spanned by the $ (2j+1)$ eigenvectors
 of $\hat{J}_z$,
$ \ket{j m} $,  is given by the Floquet operator:

 \begin{eqnarray}
\hat{F} &=& \exp[-i \omega \hat{J}_z] \ \exp[-i \mu |\hat{J}_x|] \nonumber \\
        &=& \exp[-i \omega \hat{J}_z] \ \exp[ -i \pi/2 \hat{J}_y] \ \exp[ -i \mu |{J}_z|] \ \exp[ i \pi/2 \hat{J}_y]
\end{eqnarray}

\noindent where:

$ | \hat{J}_z| = \sum_{m=-j}^j |m| \ket{j m}$

\noindent The density operator $ \hat{\rho} $ is now represented
by the corresponding Husimi function $ Q_{\rho} ( \theta, \phi) =
\bra{j \theta \phi} | \hat{\rho}  \ket{j \theta \phi} $ ( that is,
its diagonal matrix element in the basis of coherent states) and
its time evolution is given by:

\begin{equation}
 {\partial}_t \ Q_{\rho} = {\cal G} \ Q_{\rho}
\end{equation}

\noindent The Husimi propagator $ \exp [{\cal G}] $ is a unitary
matrix of dimension $ (2 j + 1 )^2$. As for the classical case
their matrix elements $ \exp[{\cal G}] =
 \exp[ {\cal G}_{R_z(\omega)}]  \ \exp[ {\cal G}_{R_y(-\pi /2)}] \ \exp[ {\cal
 G}_{|R_z(\mu)|}] \
 \exp[ {\cal G}_{R_y(\pi /2)}]$
will be calculated in the basis of the spherical harmonics.

\noindent As shown in \cite{Haa2} the Husimi propagator for rotations is identical to its classical
counterpart. We are then left with the calculation of $ \exp[ {\cal G}_{|R_z(\mu)|}])_{l m,l' m'}$, whose
spectral representation is:

\begin{equation}
\exp[ {\cal G}_{|R_z(\mu)|}])= \sum_{m_1,m_2=-j}^j Q_{\ket{j m_1}
\bra{j m_2}|} \ \exp[-i ( \mu |m_1| -|m_2|)] \ P_{\ket{j m_1}
\bra{j m_2}|} \label{pro}
\end{equation}

\noindent where $ Q $ and $ P $ ( the Q- and P-functions corresponding to $ \ket{j m_1} \bra{j m_2}|$ )
are respectively
the right-hand and left-hand eigenfunctions of $ \exp[ {\cal G}_{|R_z(\mu)|}]) $ with eigenvalue
$ \exp[-i \mu (|m_1|-|m_2|)]$.
 To express the matrix elements of the propagator ( \ref{pro} ) in the basis of the spherical harmonics, we need
 to calculate the scalar products $ <Y_{l m}|Q_{\ket{j m_1} \bra{j m_2}|}> $ and
$ <Y_{l m}|P_{\ket{j m_1} \bra{j m_2}|}> $ which are sums of Clebsch Gordan coefficients.
We finally obtain:

\begin{equation}
\exp[ {\cal G}_{|R_z(\mu)|}])_{l m,l' m'} = \delta _{m,m'}
{{2l+1}\over {2j+1}} \sum_{m_1=max(-j+m,-j)}^{min(j,j+m)} {\cal
C}_{j (m_1-m),l m}^{j m_1} \ {\cal C}_{j (m_1-m),l' m}^{j m_1}
\exp[-i \mu (|m_1| -|m_1-m|)] \label{quan}
\end{equation}

\noindent As in the classical case the matrix elements of the propagator are diagonal in $ m $ , but
non diagonal in $ l $.

We can  easily show that when  $ \exp [{\cal G}] $ is truncated to
a dimension $ N = (l_{max}+ 1 )^2 $ with $ l_{max} << 2 j $ it
approaches its classical counterpart $ {\cal P}^{(N)} $. In this
limit  the sum in eq.(\ref{quan}) can be approximated by:

\begin{equation}
\exp[ {\cal G}_{|R_z(\mu)|}])_{l m,l' m'} \approx \delta _{m,m'} {{2l+1}\over {2j+1}} [
\exp[i \mu m] \sum_{m_1=-j}^0 {\cal C}_{j (m_1-m),l m}^{j m_1} {\cal C}_{j (m_1-m),l' m}^{j m_1} +
\exp[-i \mu m] \sum_{m_1=1}^j {\cal C}_{j (m_1-m),l m}^{j m_1} {\cal C}_{j (m_1-m),l' m}^{j m_1} ]
\end{equation}

since $ |m| \leq l_{max} << 2j $. Using then the asymptotic
expression for $ l<<j $ of the Clebsch Gordan coefficients
\cite{ang}:
\begin{equation}
{\cal C}_{j (m_1-m),l m}^{j m_1} {\rightarrow} \sqrt{{4 \pi }
\over (2 l + 1)} Y_{l m}(\theta,0)
\end{equation}

\begin{figure}[h]
\begin{center}
\includegraphics*[width=6cm,angle=-90]{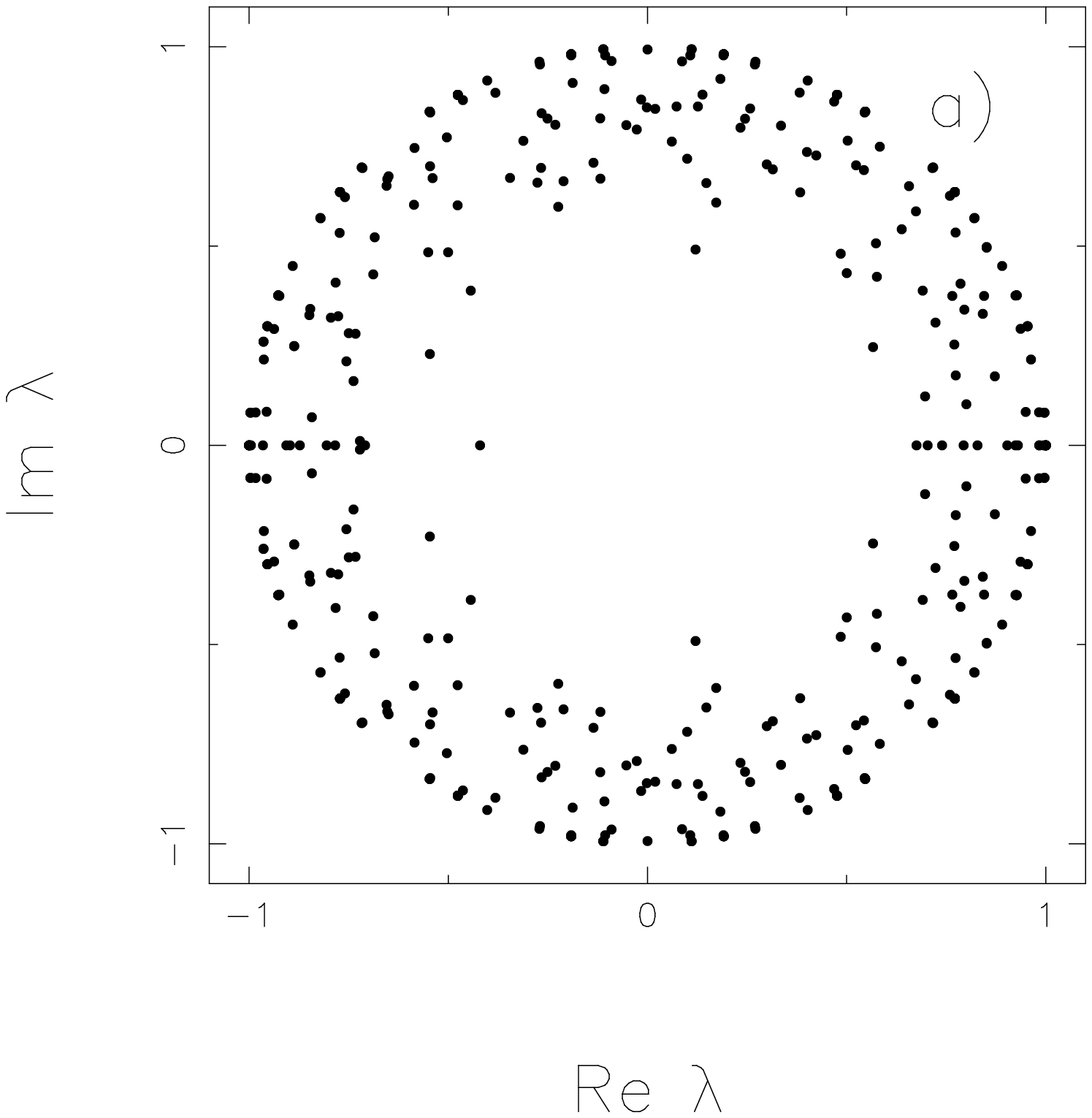}
\vspace{0.2cm}
\includegraphics*[width=6cm,angle=-90]{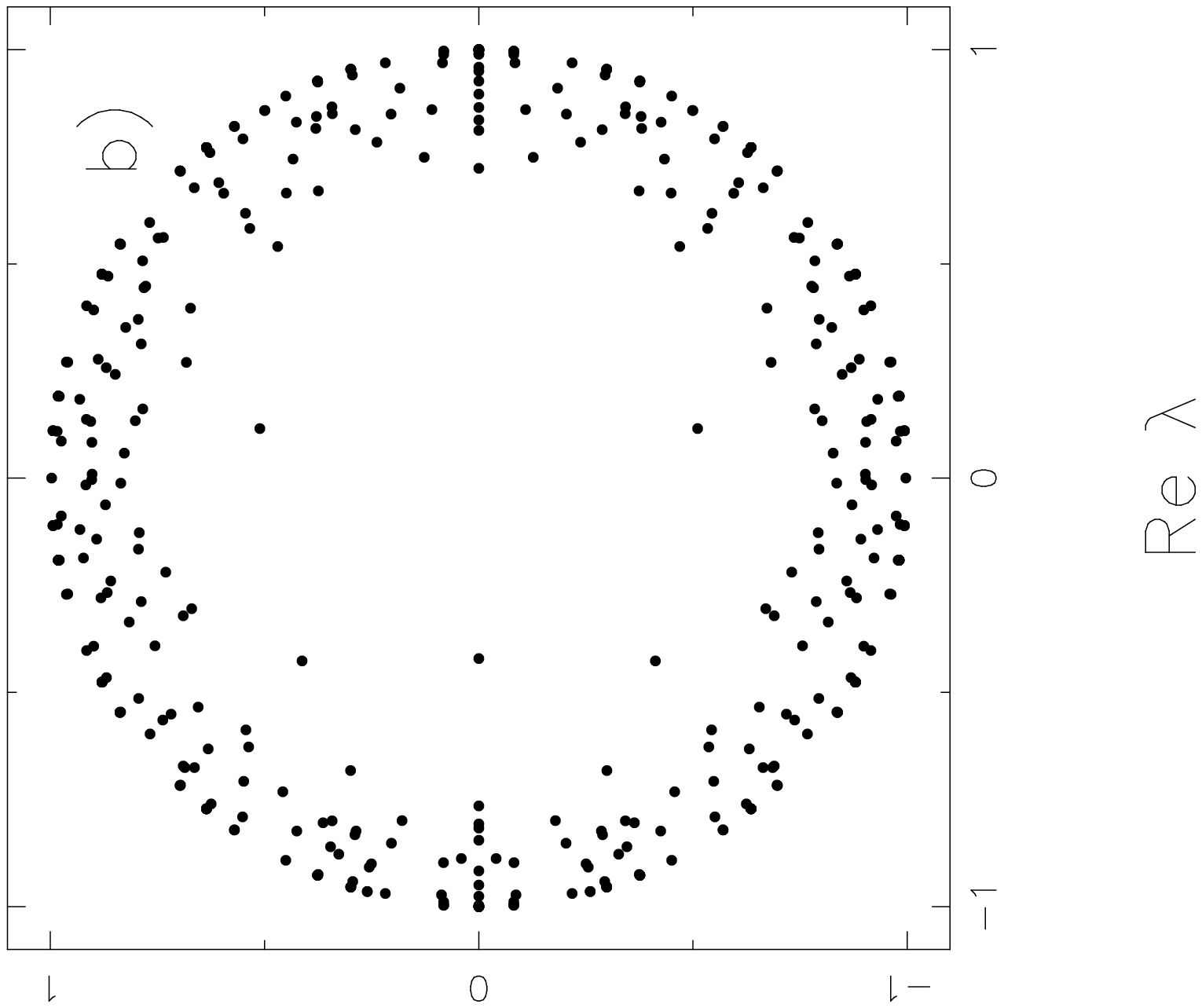}
\vspace{0.2cm}
\includegraphics*[width=6cm,angle=-90]{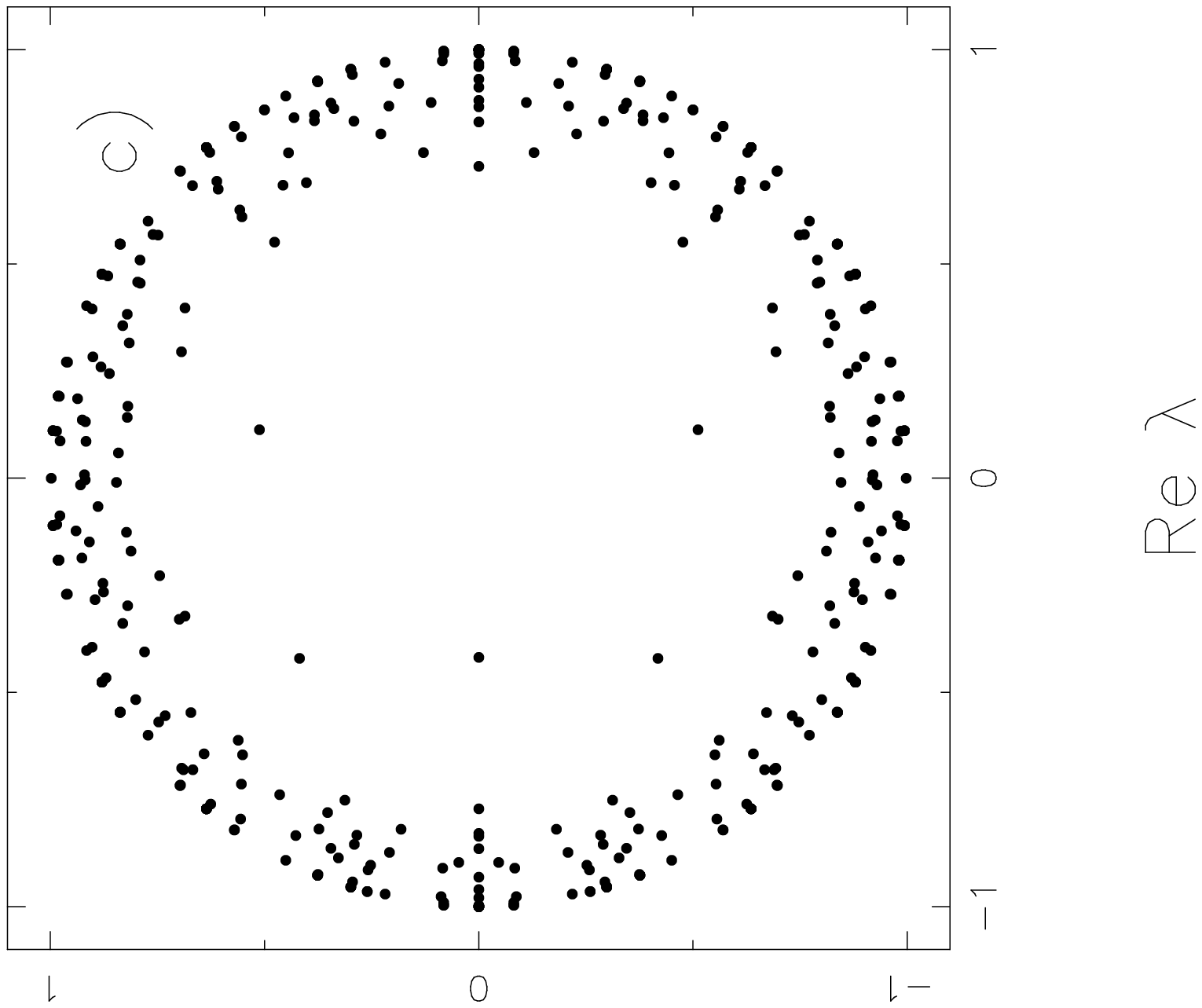}
\vspace{0.2cm}
\caption{%
Quantal spectrum (Case I) corresponding to $l_{max}=20$ for j=100,200,300.
\label{comp1} }
\end{center}
\end{figure}

\begin{figure}[h]
\begin{center}
\includegraphics*[width=6cm,angle=-90]{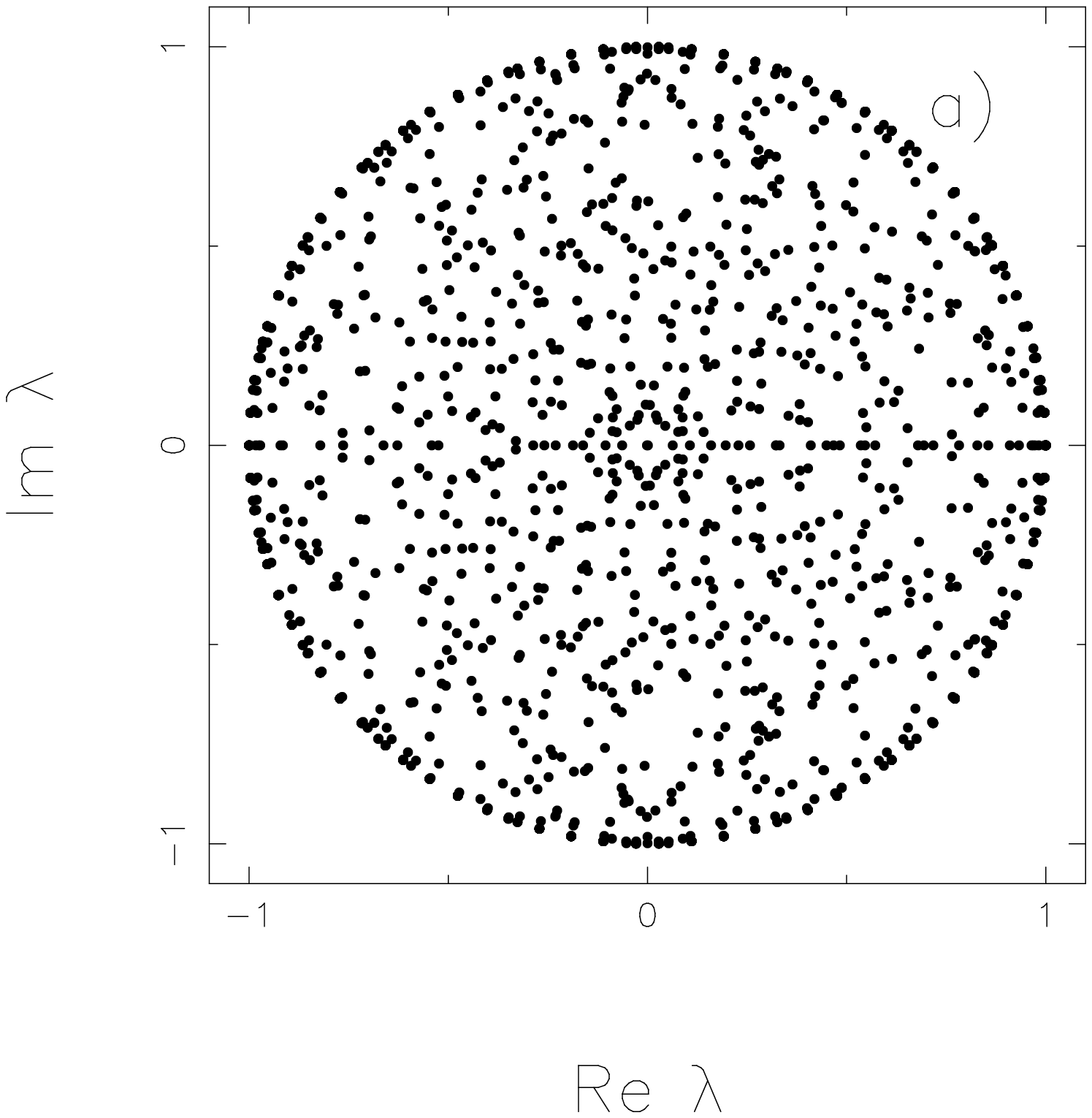}
\vspace{0.2cm}
\includegraphics*[width=6cm,angle=-90]{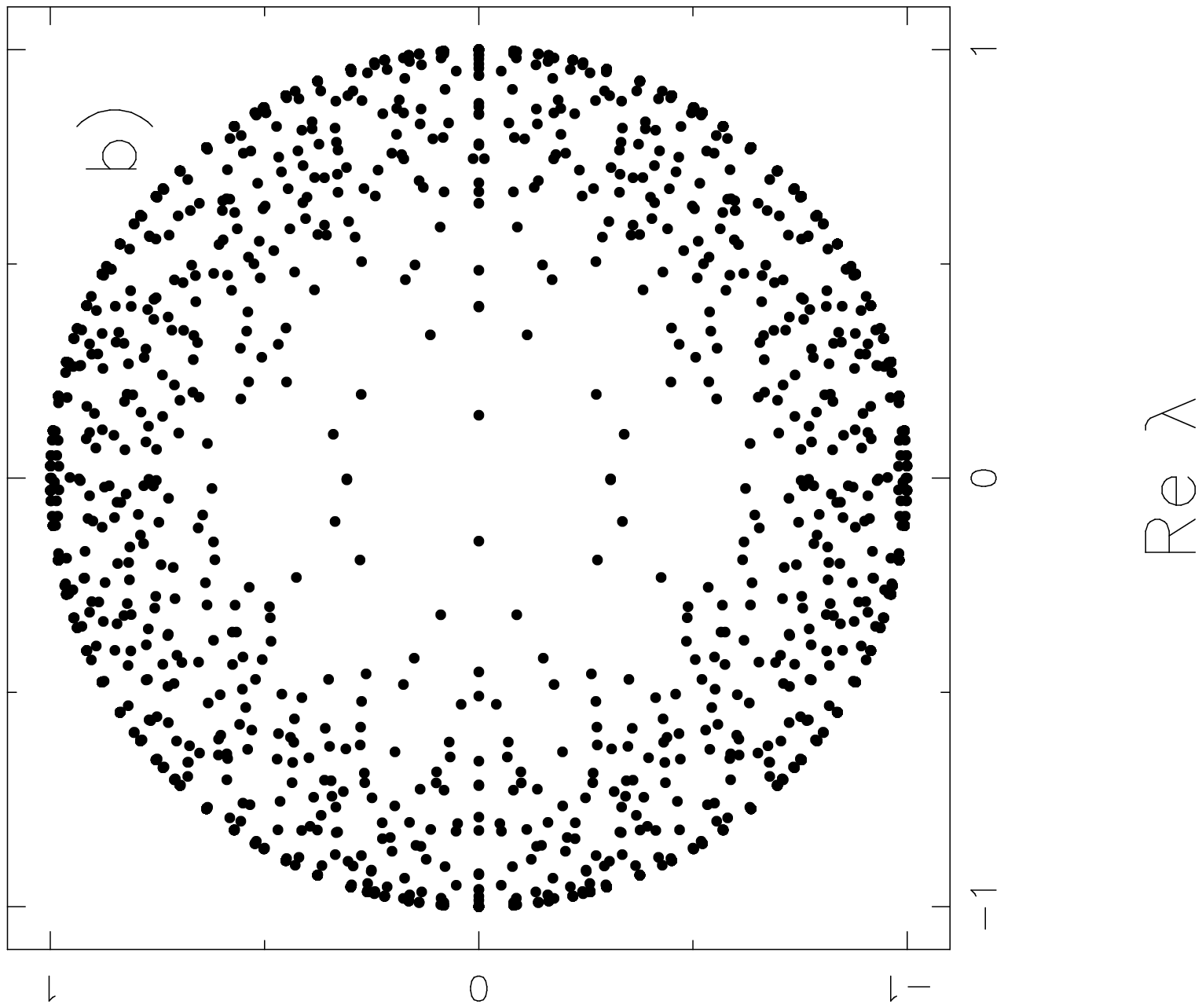}
\vspace{0.2cm}
\includegraphics*[width=6cm,angle=-90]{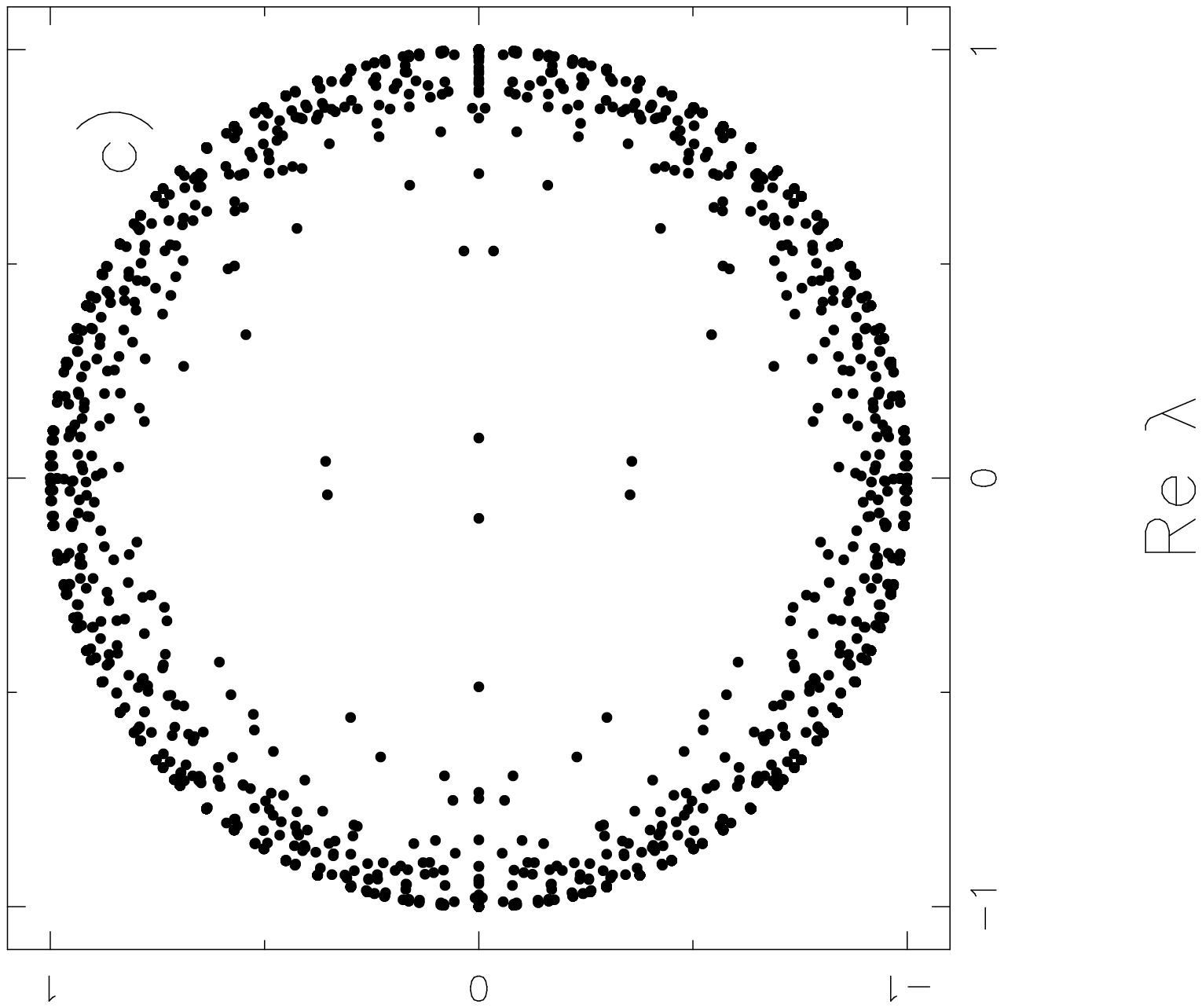}
\vspace{0.2cm}
\caption{%
Quantal spectrum (Case I) corresponding to $l_{max}=40$ for $j=100,200,700$.
\label{comp2} }
\end{center}
\end{figure}

\begin{figure}[h]
\begin{center}
\includegraphics*[width=5cm,angle=-90]{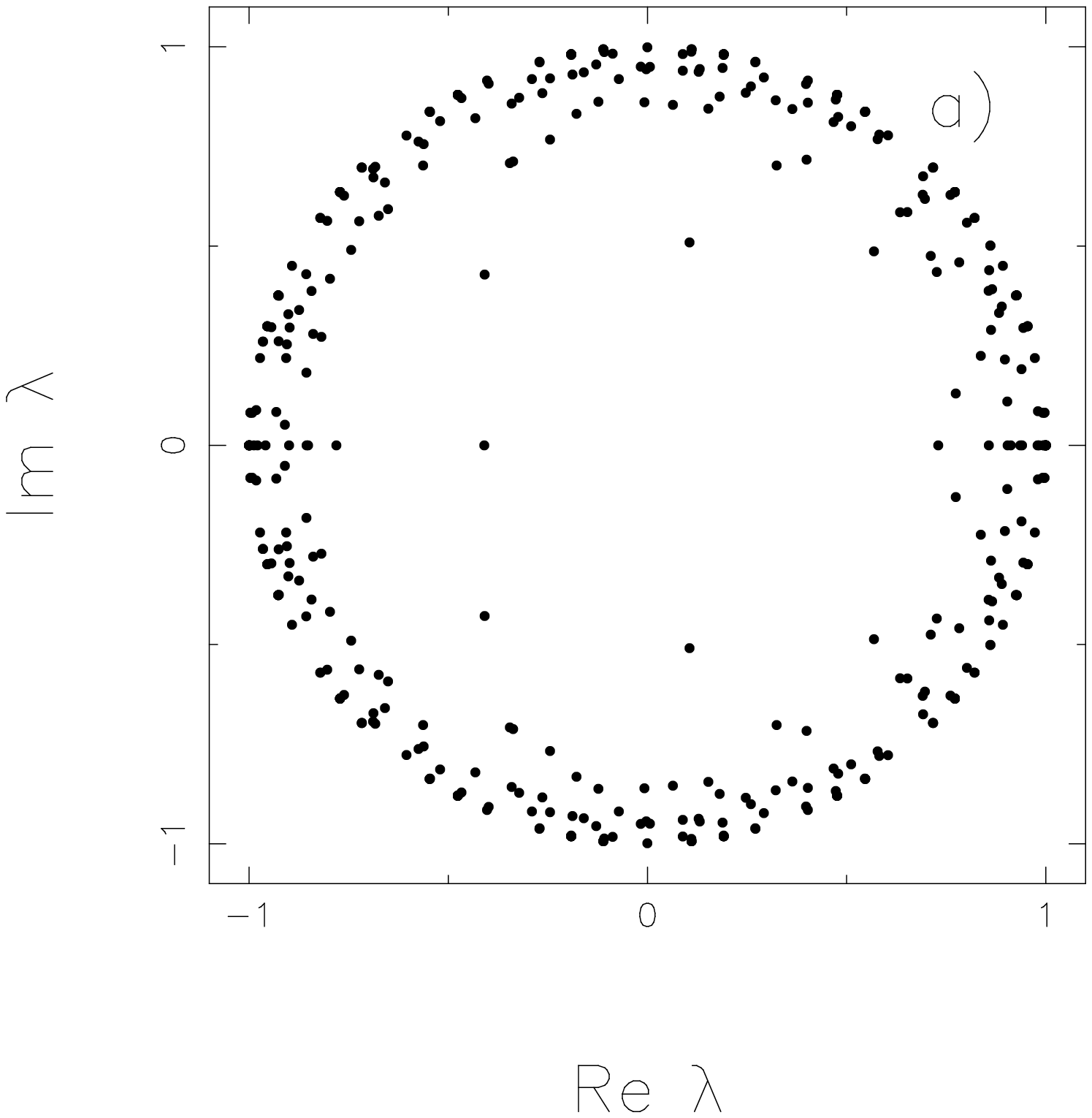}
\vspace{0.2cm}
\includegraphics*[width=5cm,angle=-90]{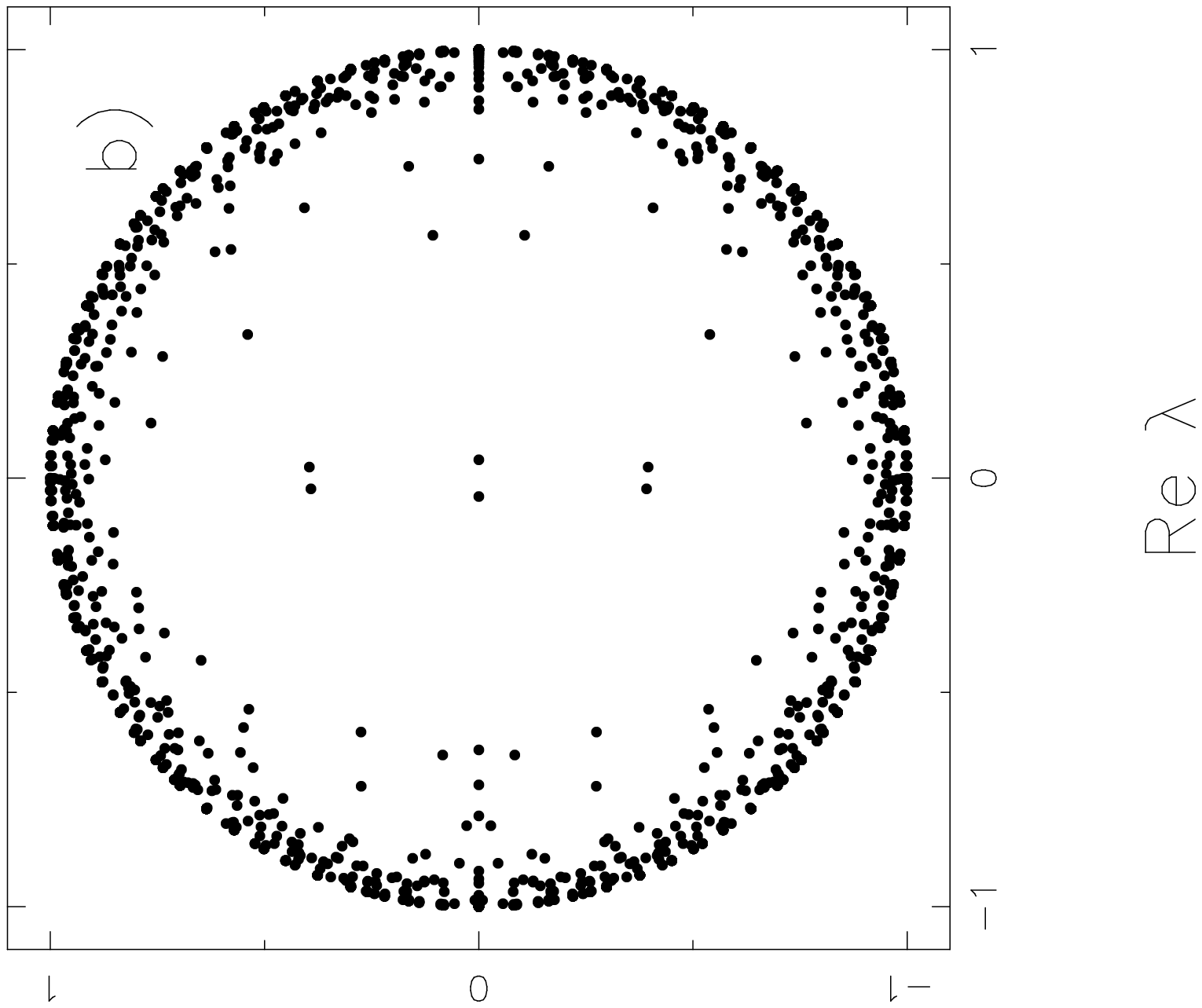}
\vspace{0.2cm}
\includegraphics*[width=5cm,angle=-90]{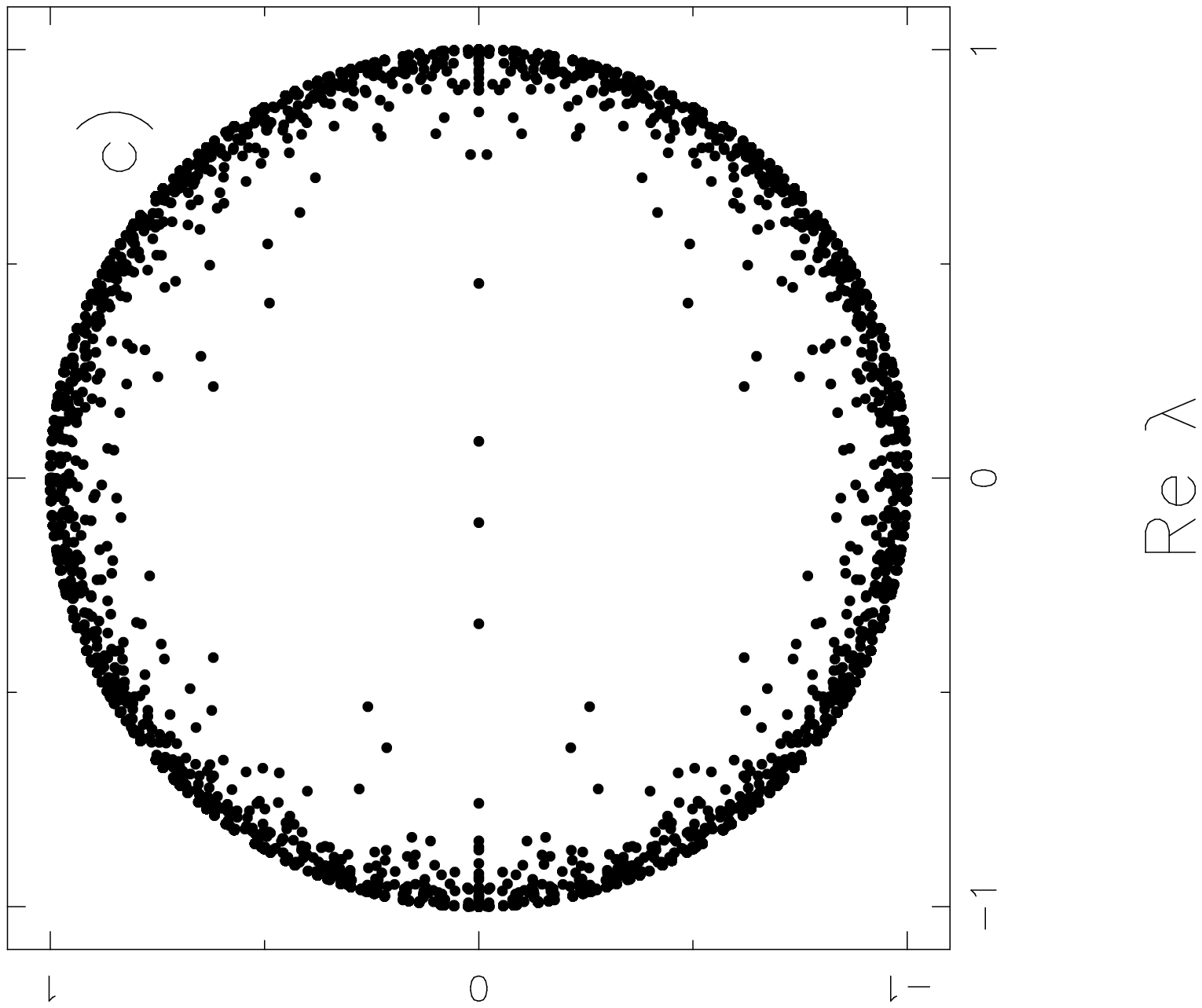}
\vspace{0.2cm}
\includegraphics*[width=5cm,angle=-90]{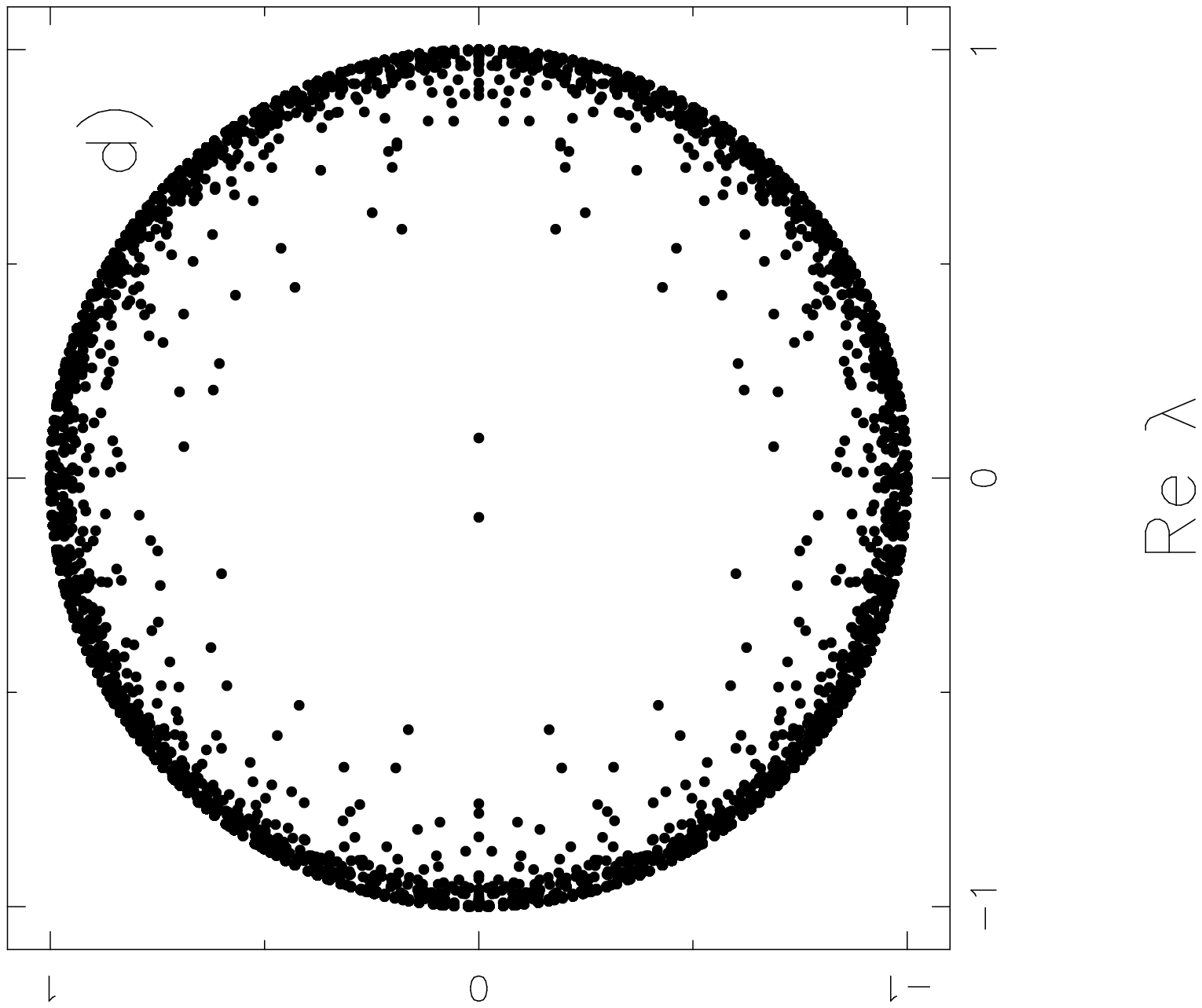}
\vspace{0.2cm}
\caption{%
Classical spectrum (Case I) for $l_{max}=20,40,60,70 $.
\label{spect1} }
\end{center}
\end{figure}

\noindent ( where $ \theta $ is the angle between $ {\hat j } $
and the z-axis)  and approximating $ {1 \over 2j+1}
\sum_{m_1=-j}^0 $ by $ \int_{-1}^0 d \cos \theta $ we recover eq
.(\ref{class3}). Therefore the matrix elements of the quantum
propagator given by eq.(\ref{quan}) coincide with their classical
counterpart of eq .(\ref{class3}) for low resolution an in this
limit it will be equivalent to study the spectrum of the FP or the
Husimi propagator. However the convergence of the quantum
mechanical propagator to the FP propagator is much slower than the
 one found for the kicked top in \cite{Haa2}, where the quantum mechanical corrections were of the order of
 $ l/2j+1$.
 This can be appreciated when comparing the quantal spectra of Fig. \ref{comp1}, with $l_{max}=20$ and
different values of $ j $, with the corresponding classical
spectrum of Fig.\ref{spect1} a). The convergence gets worse as
$l_{max}$ increases, as shown in Fig. \ref{comp2} which displays
quantal spectra for $l_{max}=40$ and different
values of $ j $ ( to be compared with Fig. \ref{spect1} b)).\\

We now diagonalize the truncated FP operator $ {\cal P}^{(N)} $,
whose elements are given by
eqs.(\ref{class1},\ref{class2},\ref{class3}) in the subspace
spanned by $ N=(l_{max}+1)^2 $ spherical harmonics with $ 0  \leq
l \leq  l_{max} $. This corresponds to a resolution of phase space
structures of area $ 4 \pi /((l_{max}+1)^2 $. Two sets of
parameters $ \omega $, $ \mu $ are considered in the map
eq.(\ref{map1}) with the corresponding phase portraits shown in
Fig.\ref{spect0}. In the first case ( Case I, $ \omega=  \mu= \pi
(\sqrt{5}-1)$) few big elliptic islands are present, corresponding
to orbits of low periodicity. In the second ( Case II, $ \omega=
1,$ $ \mu=0.05 $ ) the quasi tori surviving in correspondence to
the integrable case ( $\mu = 0$) are visible.

In Fig. \ref{spect1}  we display the eigenvalue spectrum of $
{\cal P}^{(N)} $ corresponding to Case I for different values of $
l_{max} $. For coarse resolution $ l_{max}=20 $ most of the
eigenvalues $ \lambda_i $  concentrate inside a ring corresponding
to $ 0.9 \leq |\lambda_i| \leq 1 $. As the resolution increases,
the outer ring becomes narrower ( for $ l_{max}=70 $, $ 0.95 \leq
|\lambda_i| \leq 1 $) and the fraction of the eigenvalues located
in the inner disk decreases. In Case II, the situation is very
much the same for $ l_{max}=70 $, that is, there is a very high
concentration of eigenvalues inside a narrow ring and a few values
lying in the inner disk. However for low resolution  $ l_{max}=20
$ the distribution looks different: most of the eigenvalues lie
close to the unit circle. Other spectra have been computed for
different values of the map`s parameters $ \omega$ and  $ \mu $,
corresponding to different types of phase space portraits with
more or less dominant islands. The characteristics of the spectra
differ from case to case depending on the number of domains that
can be resolved at each given resolution. This non universality
was also remarked in \cite{Spi2} where the calculation of quantal
spectra of triangular billiards on the sphere showed that the
level distributions do not follow a universal statistics but can
rather be understood in reference to the phase space portrait
corresponding to each particular case. However the general
features of spectra obtained with high resolution are the same in
all cases. There is always a densely populated outer ring,
corresponding to unimodular and almost unimodular eigenvalues, and
a small fraction of the eigenvalues lying in the inner disk. The
eigenvalues move about when increasing $ l_{max} $ and no 'frozen'
eigenvalues can be individualized.

\begin{figure}[h]
\begin{center}
\includegraphics*[width=8cm,angle=-90]{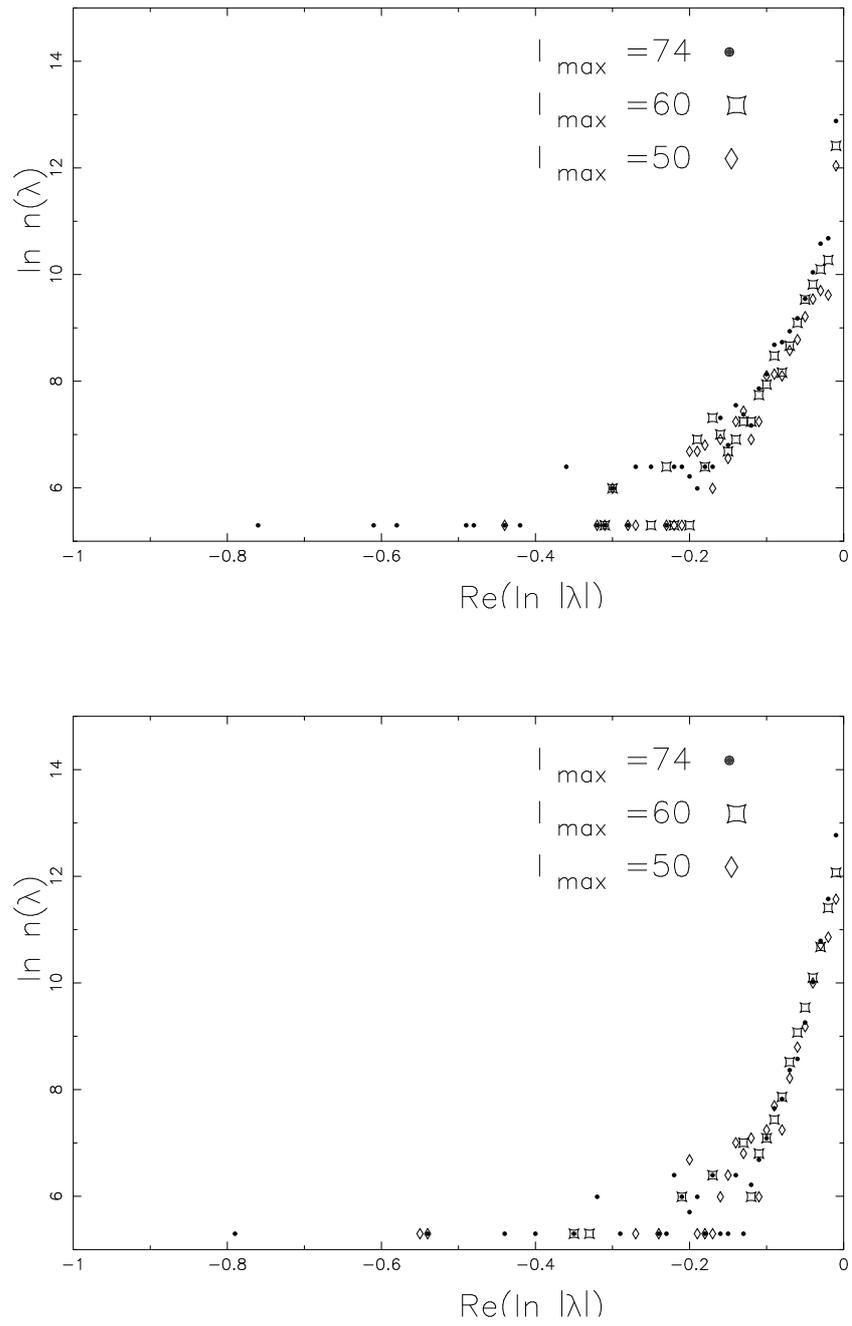}
\vspace{0.2cm}
\caption{%
Number of eigenvalues (in a logarithmic scale) as a function of $ln |\lambda_i|$
for different values of $l_{max}$ in a) Case I and b) Case II.
\label{profi} }
\end{center}
\end{figure}

\begin{figure}[h]
\begin{center}
\includegraphics*[width=6cm,angle=-90]{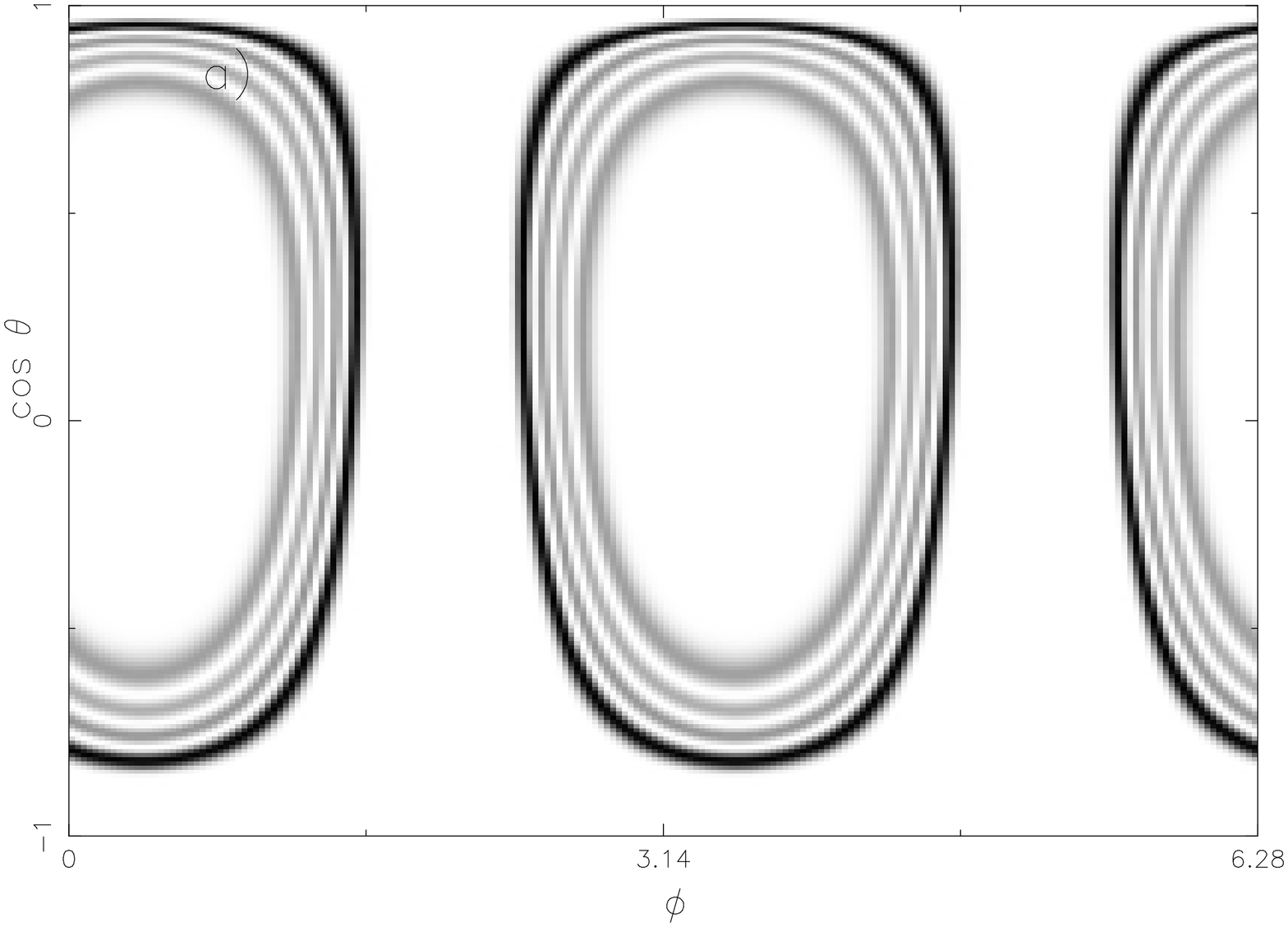}
\vspace{0.2cm}
\includegraphics*[width=6cm,angle=-90]{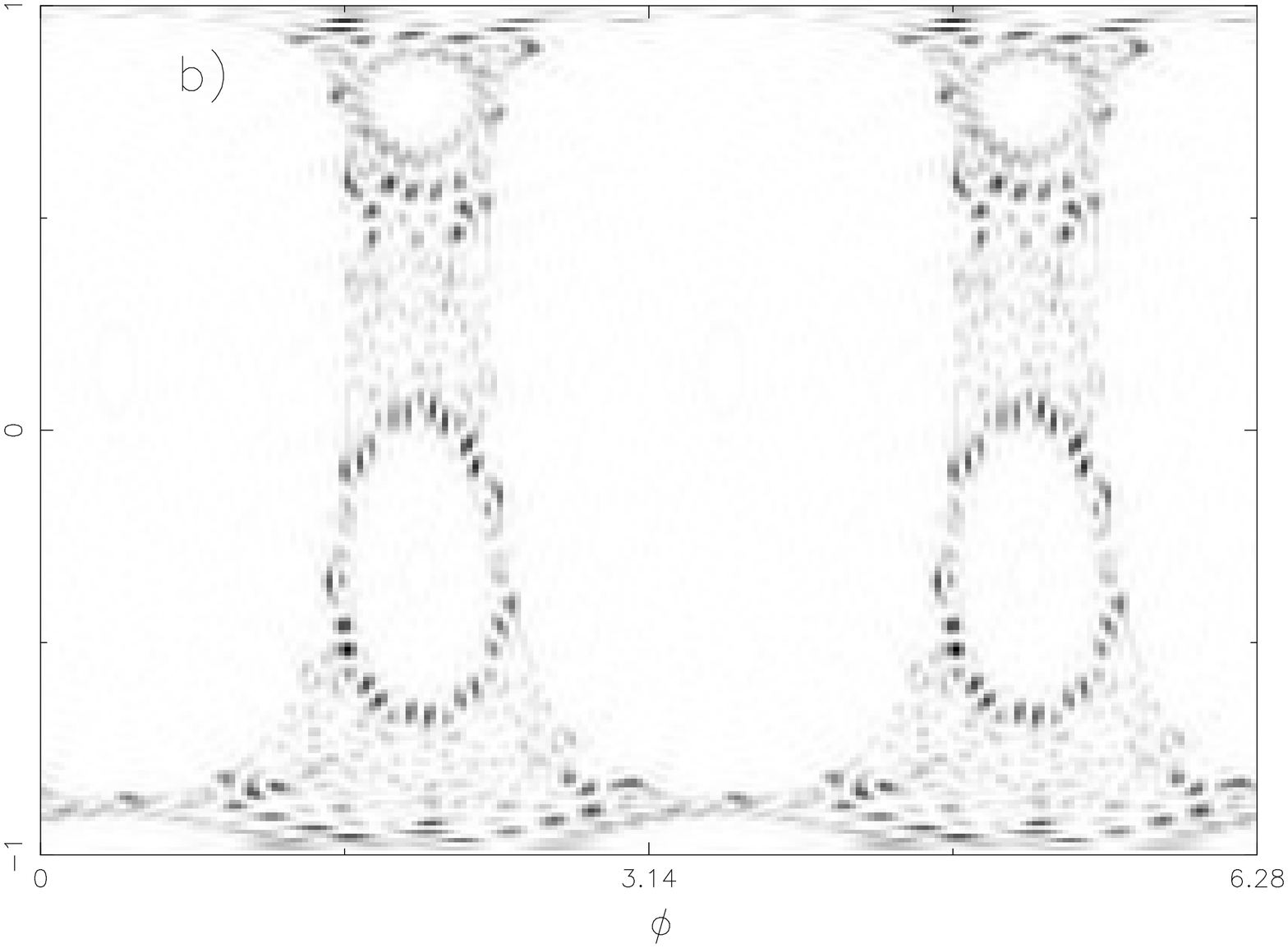}
\vspace{0.2cm}
\includegraphics*[width=6cm,angle=-90]{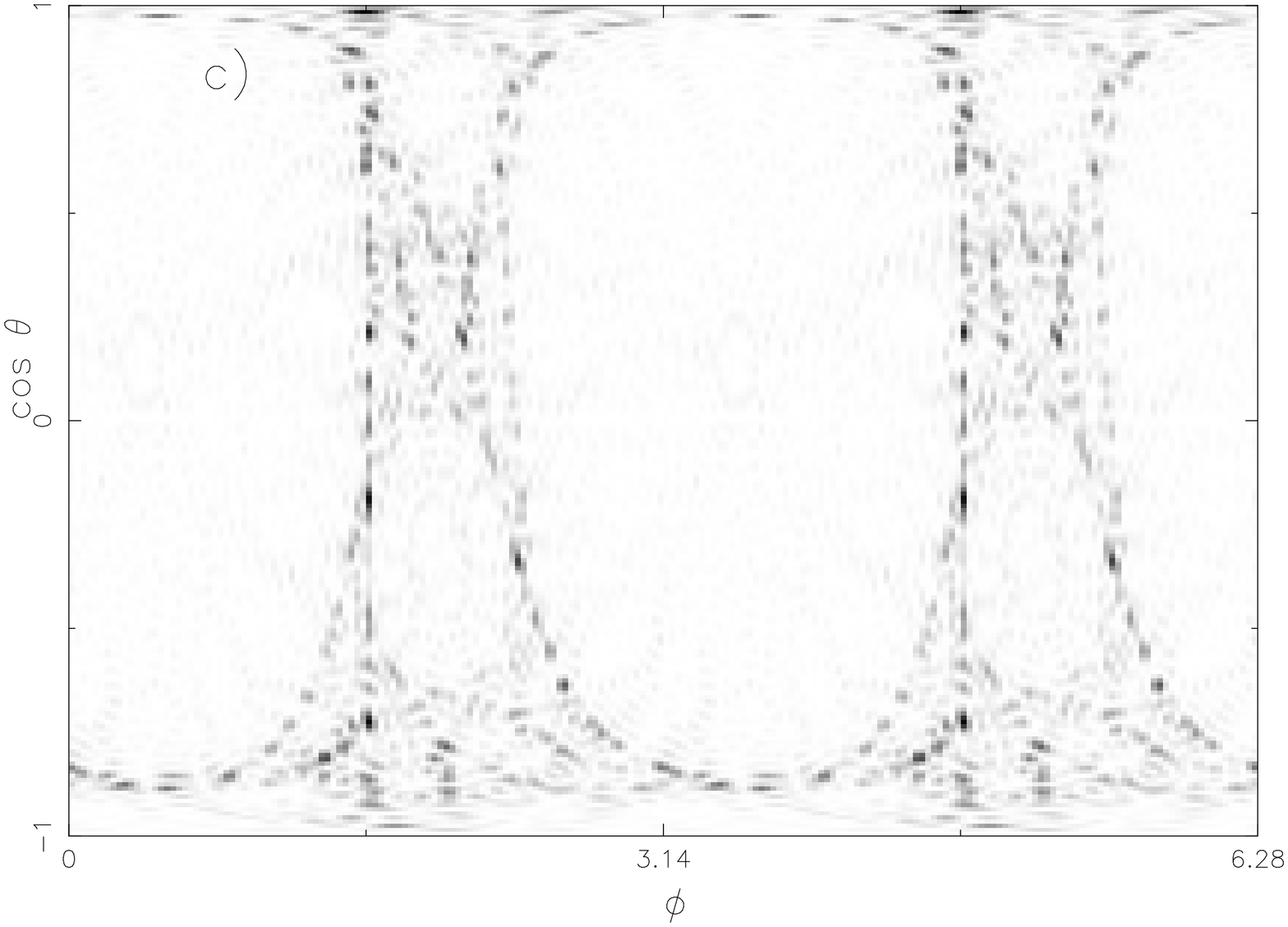}
\vspace{0.2cm}
\includegraphics*[width=6cm,angle=-90]{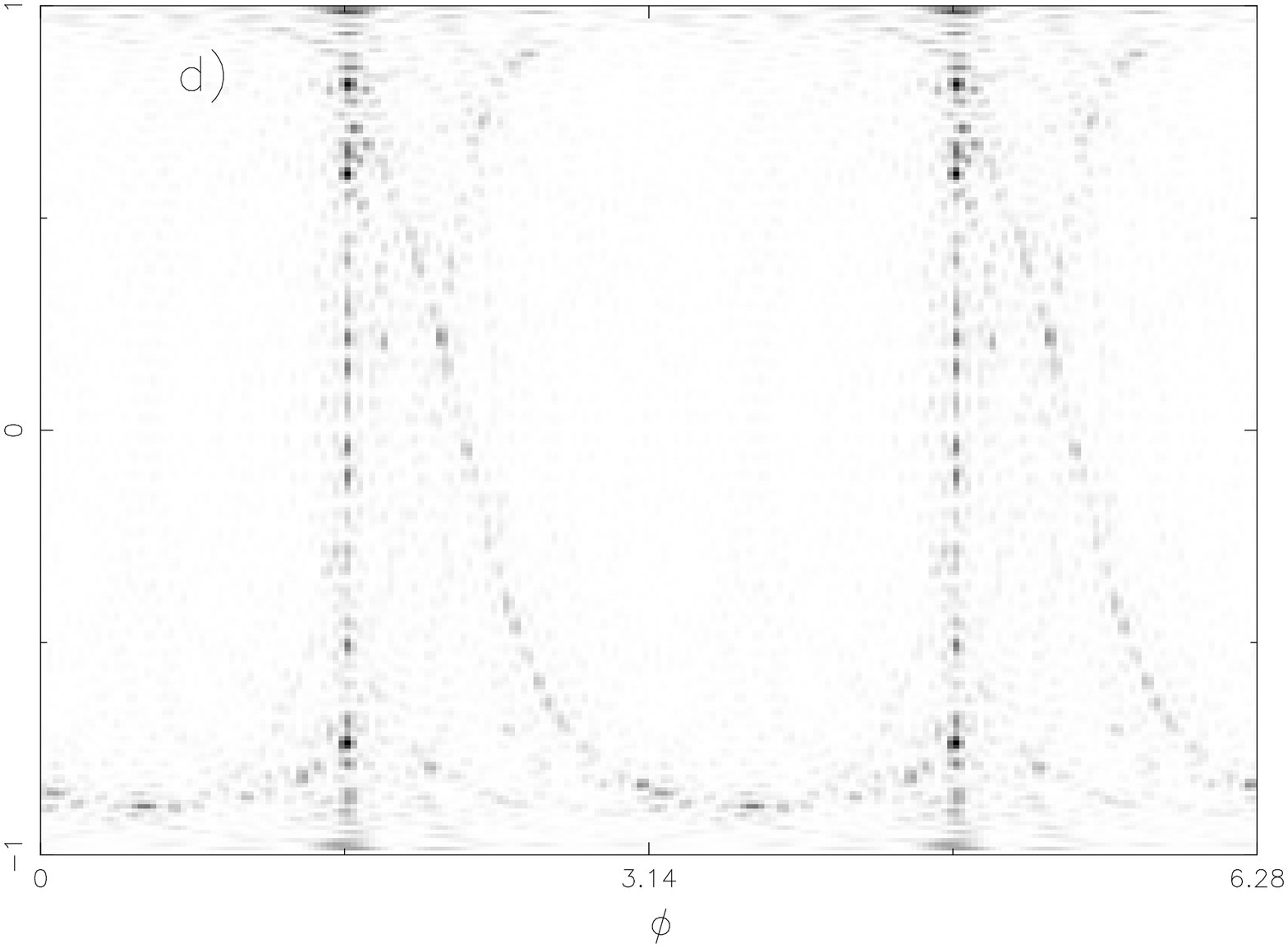}
\vspace{0.2cm}
\caption{%
Eigenfunctions of the classical propagator $ {\cal P}^{(N)} $ (
Case I , $l_{max}=70 $)  corresponding to eigenvalues with a)$
|\lambda|=1 $, b)$ |\lambda|=0.97 $, c)$ |\lambda|=0.85 $, d)$
|\lambda|=0.72 $. \label{spect3} }
\end{center}
\end{figure}

\begin{figure}[h]
\begin{center}
\includegraphics*[width=6cm,angle=-90]{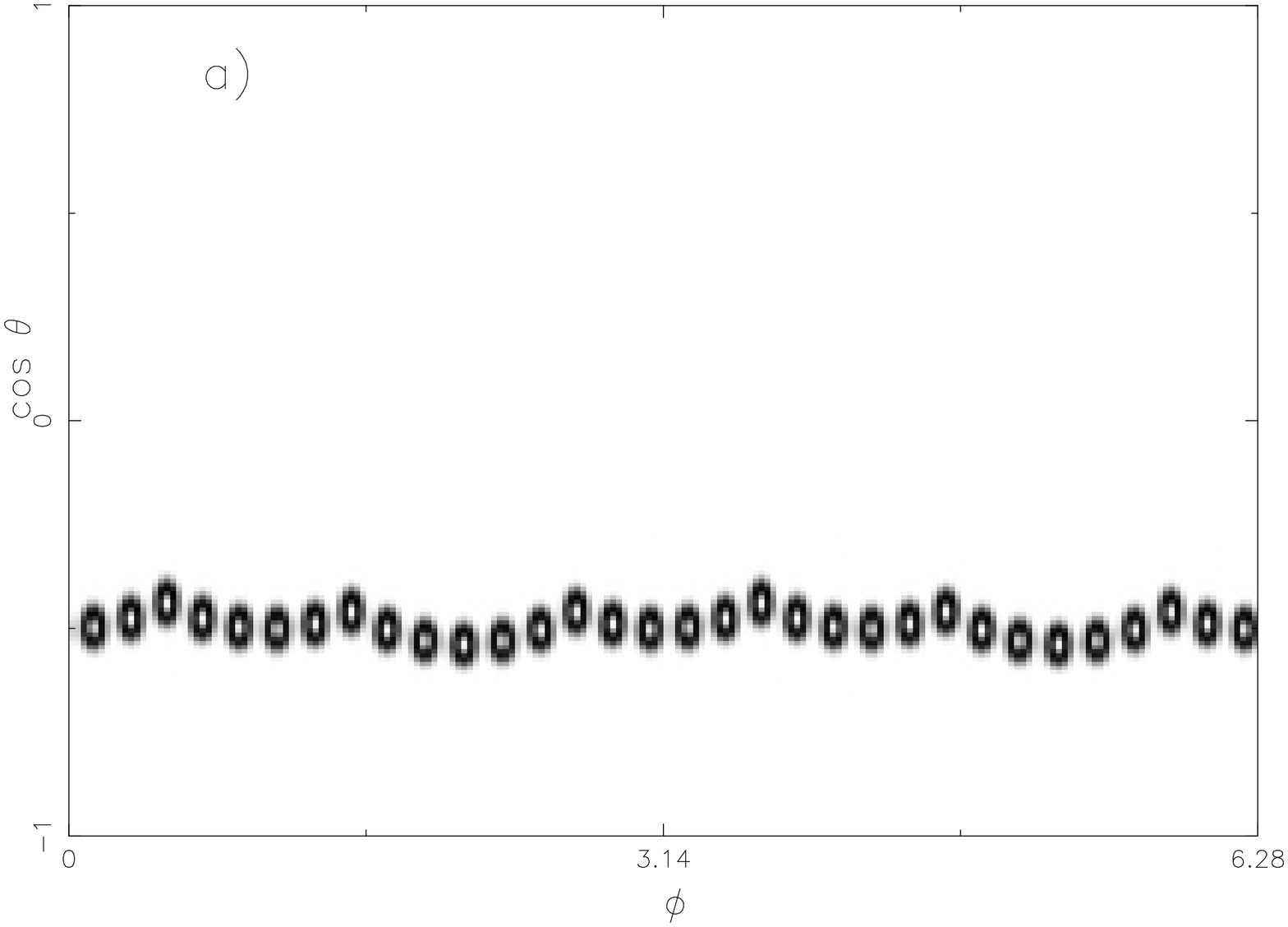}
\vspace{0.2cm}
\includegraphics*[width=6cm,angle=-90]{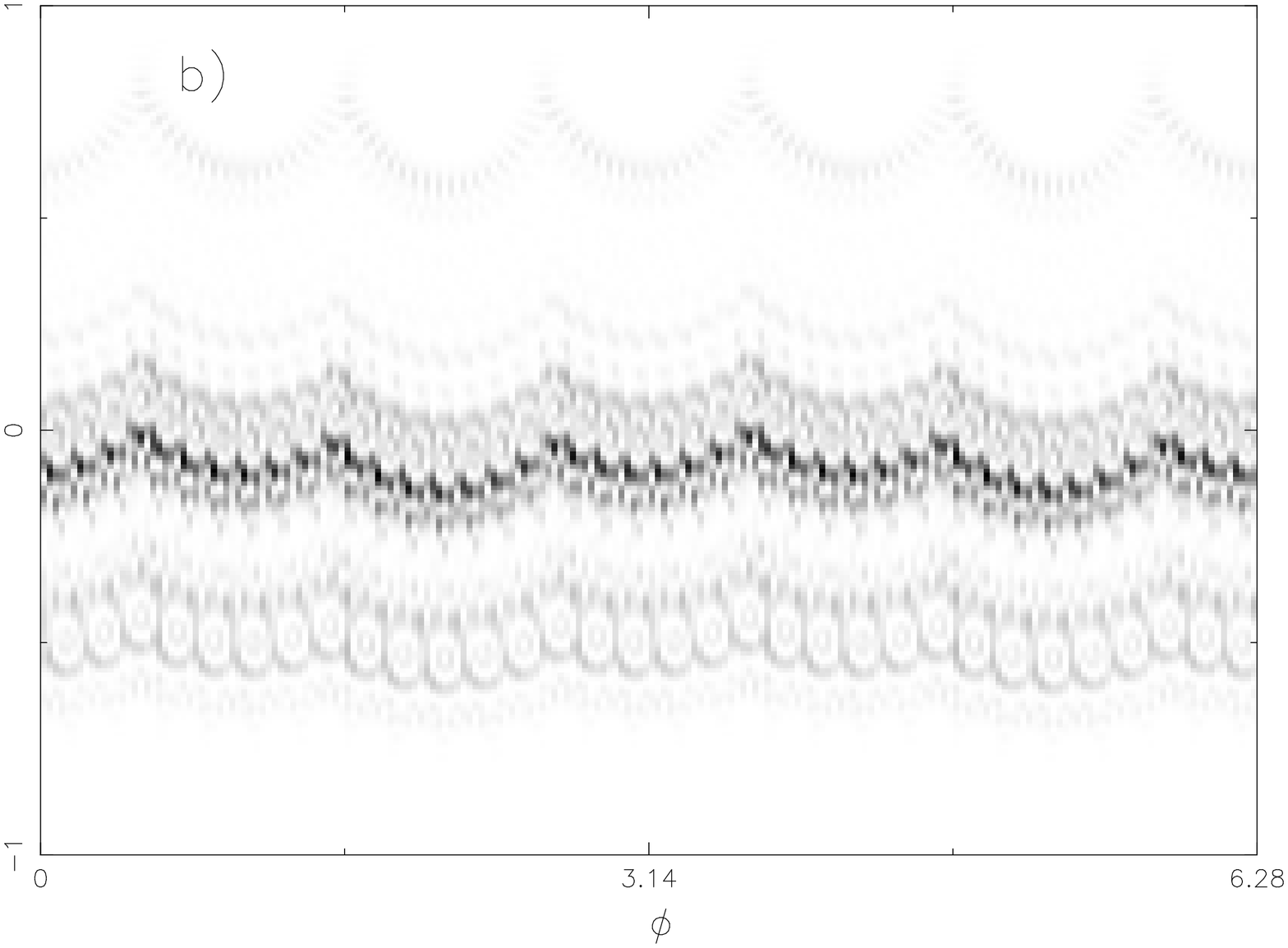}
\vspace{0.2cm}
\includegraphics*[width=6cm,angle=-90]{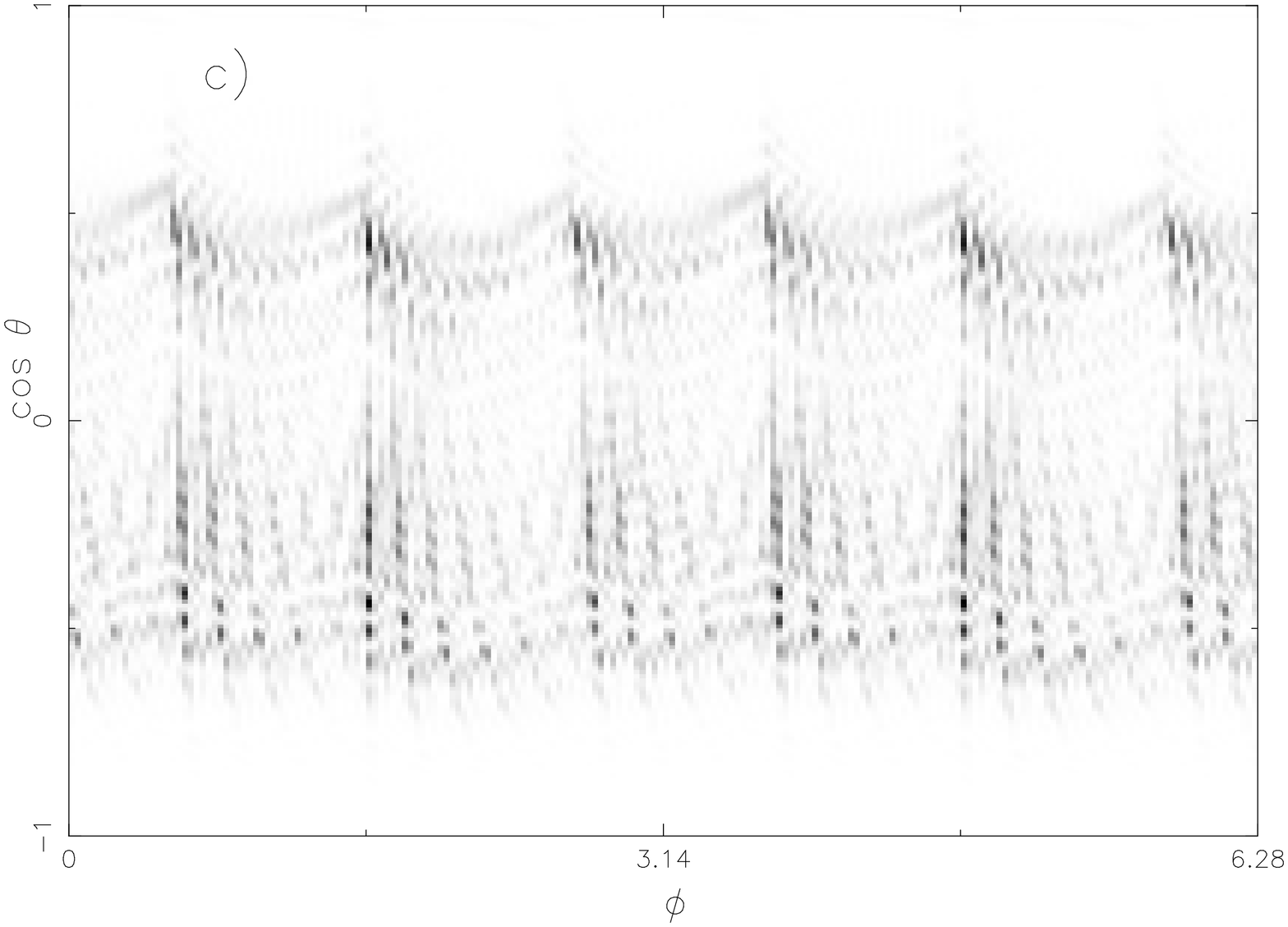}
\vspace{0.2cm}
\includegraphics*[width=6cm,angle=-90]{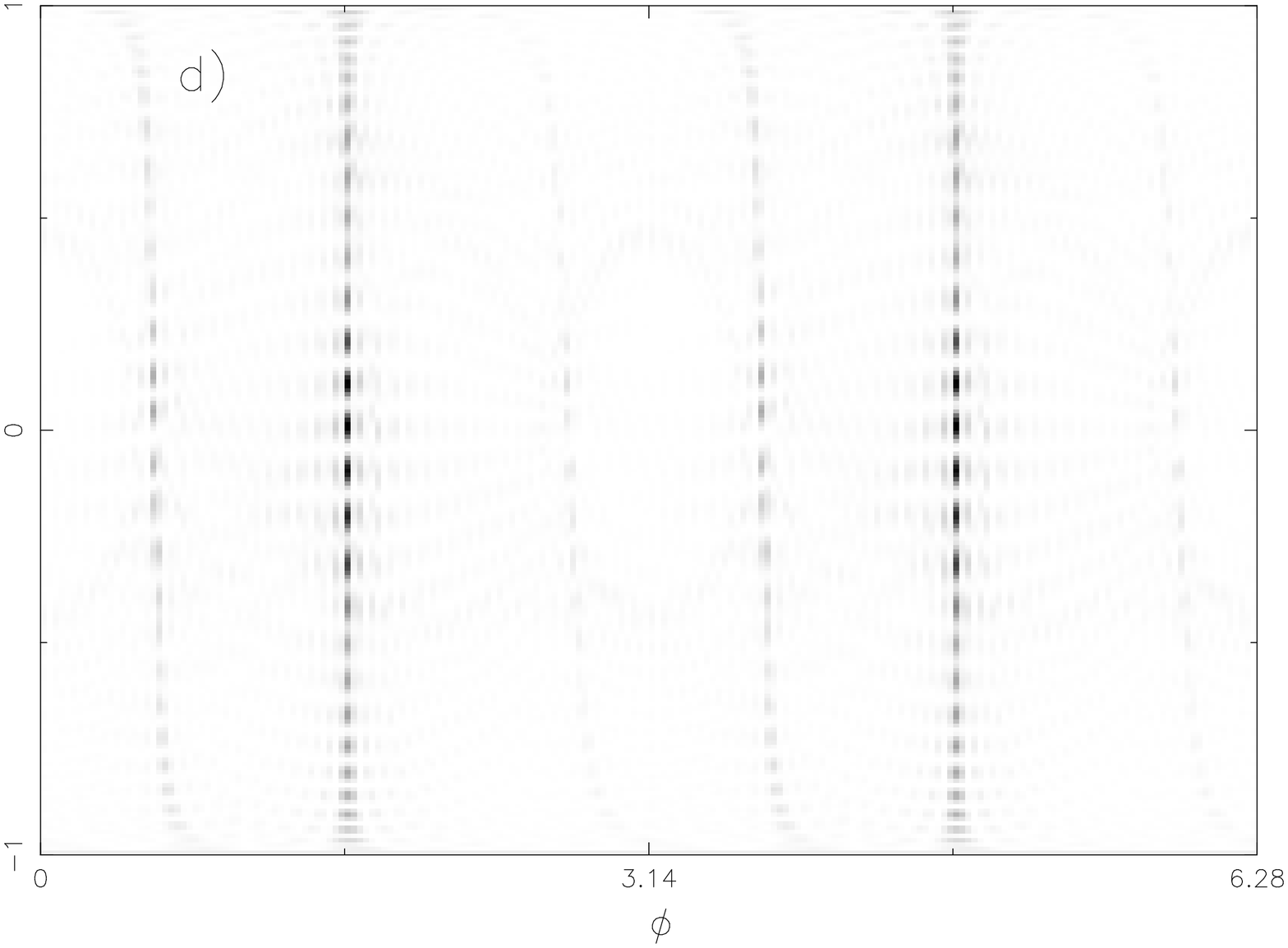}
\vspace{0.2cm}
\caption{%
Eigenfunctions of the classical propagator $ {\cal P}^{(N)} $ (
Case II , $l_{max}=70 $) corresponding to eigenvalues with a)$
|\lambda|=1 $, b)$ |\lambda|=0.95 $, c)$ |\lambda|=0.88 $, d)$
|\lambda|=0.75 $. \label{spect4} }
\end{center}
\end{figure}

A better understanding of the spectrum is achieved by plotting the distributions of the $N$ eigenvalues
as a function of their modulus for different resolutions.
As shown in Figs.\ref{profi} (for Case I and II, and $l_{max}=50,60,74$)
 we can distinguish three regions: a peak at $ 0.95 \leq |\lambda_i| \leq 1 $ , an
intermediate domain corresponding to approximately $ 0.85 \leq
|\lambda_i| \leq 0.95 $ which looks very much the same for all
resolutions and a small number of values with $|\lambda_i| \leq
0.85 $. While the peak corresponding to $ 0.95 \leq |\lambda_i|
\leq 1 $  grows for increasing values of $ l_{max} $, the number
of states with $ |\lambda_i| \leq 0.95 $ remains fairly constant,
so that the fraction of non unimodular eigenvalues decreases with
increasing resolution. Moreover, the numerical results seem to
imply in the intermediate region a relationship $ n(|\lambda|)
\sim |\lambda|^{\alpha}$ where the exponent $ \alpha $ is
independent of the resolution. At this stage we cannot interpret
this observation. The eigenfunctions corresponding to these three
regions are of different nature. Some examples are represented
 in Fig.\ref{spect3} for Case I and $ l_{max}=70 $ and in Fig.\ref{spect4} for Case II and $ l_{max}=70 $.
Eigenfunctions with unimodular eigenvalue are shown in Fig. \ref{spect3}a) and Fig.\ref{spect4}a):
they are supported by the elliptic islands that can be resolved at the corresponding $ l_{max} $.
Eigenfunctions corresponding to  quasi unimodular eigenvalues ( $ 0.9 \leq |\lambda_i| \leq 0.95 $) have the
structure displayed in Fig.\ref{spect3}b) and Fig.\ref{spect4}b) . Their amplitude is
 evenly distributed over a large
number of islands.
Finally  for $ |\lambda_i| \leq 0.85 $ the eigenfunctions
are of the type shown in Fig.\ref{spect3}c),d) and Fig.\ref{spect4}c),d): they locate along
the unstabiliy line and along curves which are their first iterations by the map $ M $.

On the basis of these numerical results we can describe
the $ {\cal P}^{(N)} $ spectrum, consisting of unimodular, almost unimodular and non unimodular eigenvalues,
 and its evolution with $ N $ as follows.
Unimodular eigenvalues correspond to eigenfunctions which locate in the elliptic islands visible at the
given resolution: when $N$ grows more and more elliptic structures are resolved and their number increases.
The group of almost unimodular eigenvalues correspond to
eigenfunctions which are not fully localized
yet and spread over several islands. The fact that their number remains approximately constant with $N$, showing
a compensation between the migration of existing eigenvalues to the unit circle and the appearance of new ones,
related to still unlocalized states,
 can be explained by the existence of an infinite hierarchy of phase space
structures which are gradually resolved.
Finally, we saw that eigenvalues with moduli $|\lambda_i| \leq 0.85 $ correspond to eigenstates sharply localized on
the unstability line. These states are present for any value of $N$. However, their
number seems to be approximately constant, so that they will represent a vanishing
fraction of the total spectrum in the limit $ N \rightarrow \infty $.

As expected, the observed evolution of the $ {\cal P}^{(N)} $
spectrum with $ N $ is different from the one obtained for the
integrable system (corresponding tho $ \mu = 0 $ in our map) ,
where all eigenvalues migrate to the unit circle in the $ N
\rightarrow \infty $ limit and unitarity is recovered. The
situation is also different from the one observed for chaotic
systems, where $ {\cal P}^{(N)} $ has an essential spectrum inside
a circle of radius ($ r > 0 $) and a point spectrum inside a disk
$ r < |\lambda| < 1 $, with some eigenvalues (that can be related
to the RP resonances) persisting in their positions for increasing
values of $N$ \cite{Haa1}. In the system under study, non
unimodular eigenvalues exist and can be interpreted, as in the
chaotic case, as an effective dissipation due to the loss of
probability to finer unresolved structures. However 'freezing'
which in the case of chaotic dynamics is attributed to the self
similarity in phase space, is not observed. Thus the 'resonances'
will not be a characteristic of the classical system but rather
will depend on the resolution. In addition, the fraction of non
unimodular eigenvalues vanishes in the limit $ N \rightarrow
\infty $, since it is related to the measure of the unstable set.

Summarizing, we computed the matrix elements of the FP operator $
{\cal P}^{(N)} $ and of the quantum Husimi propagator for
discontinous piecewise linear systems and showed that they
coincide in the limit of low resolution. We investigated the
evolution of the spectra of the truncated propagator $ {\cal
P}^{(N)} $ with $N$. We observed significant differences with
respect to the results obtained for systems with chaotic or mixed
phase spaces, where this truncation procedure is used to get the
RP resonances related to the decay rate of correlations. One is
the absence in our case of stable eigenvalues persisting in their
positions for $ N \rightarrow \infty $.  Another fundamental
difference is the fact that the fraction of the non unimodular
eigenvalues, present for all values of the resolution and
corresponding to eigenfunctions localized on the unstability line
and their images , decreases for increasing values of the
resolution. This is can be taken as an indication of the zero
measure of the unstable set.

\pagebreak

\pagebreak

\pagebreak

\begin{figure}[h]
\begin{center}
\includegraphics*[width=5cm,angle=-90]{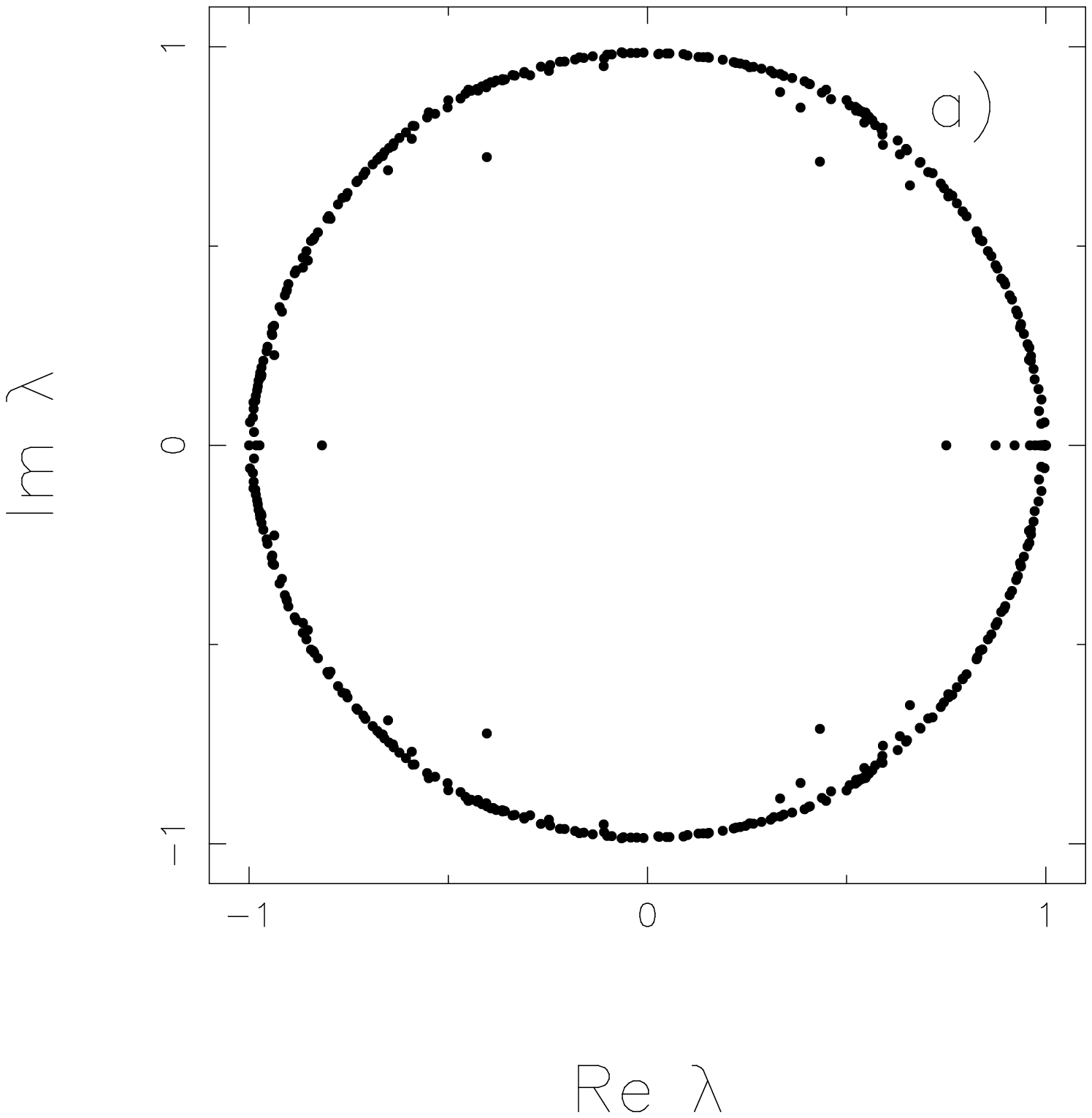}
\vspace{0.2cm}
\includegraphics*[width=5cm,angle=-90]{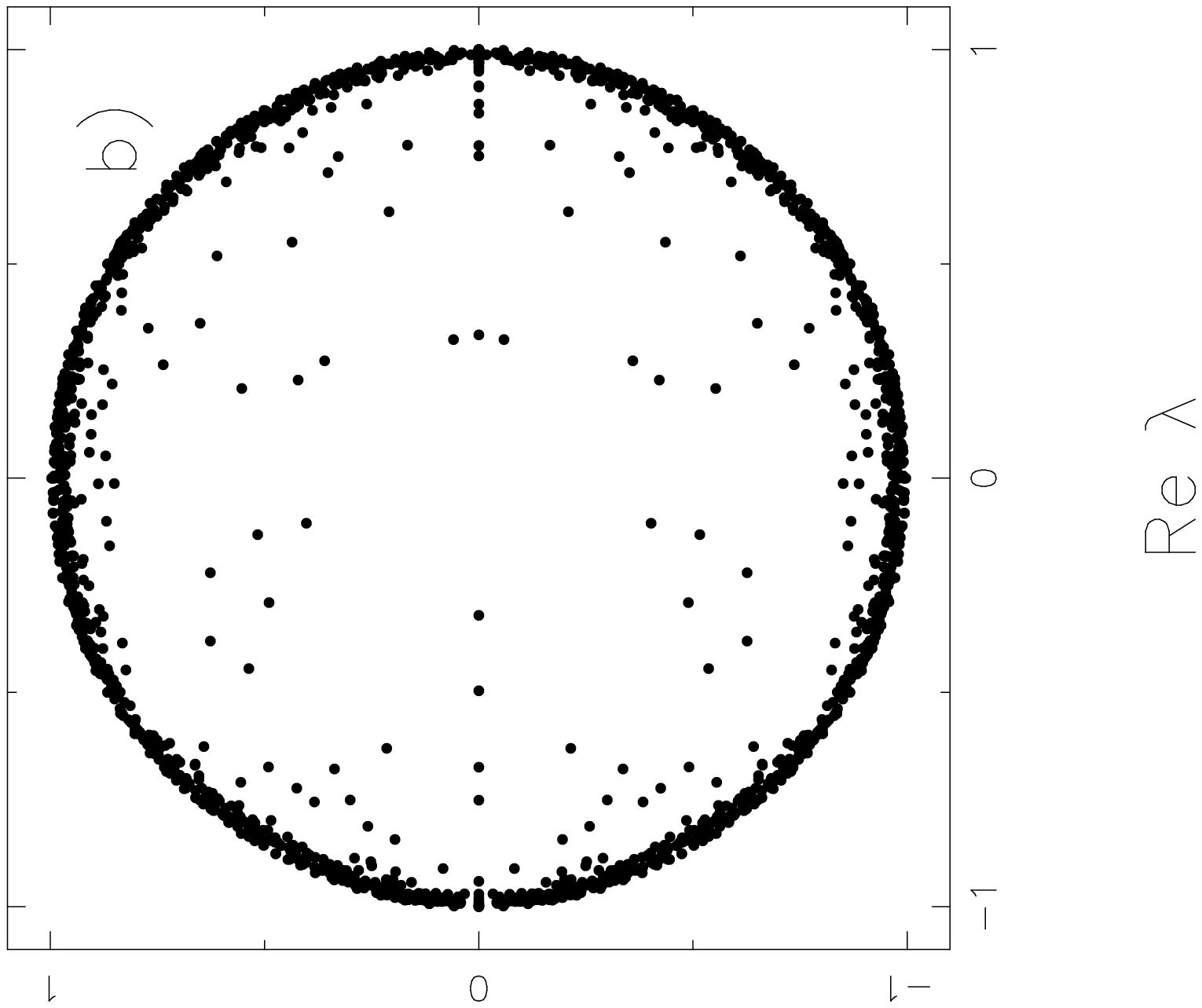}
\vspace{0.2cm}
\includegraphics*[width=5cm,angle=-90]{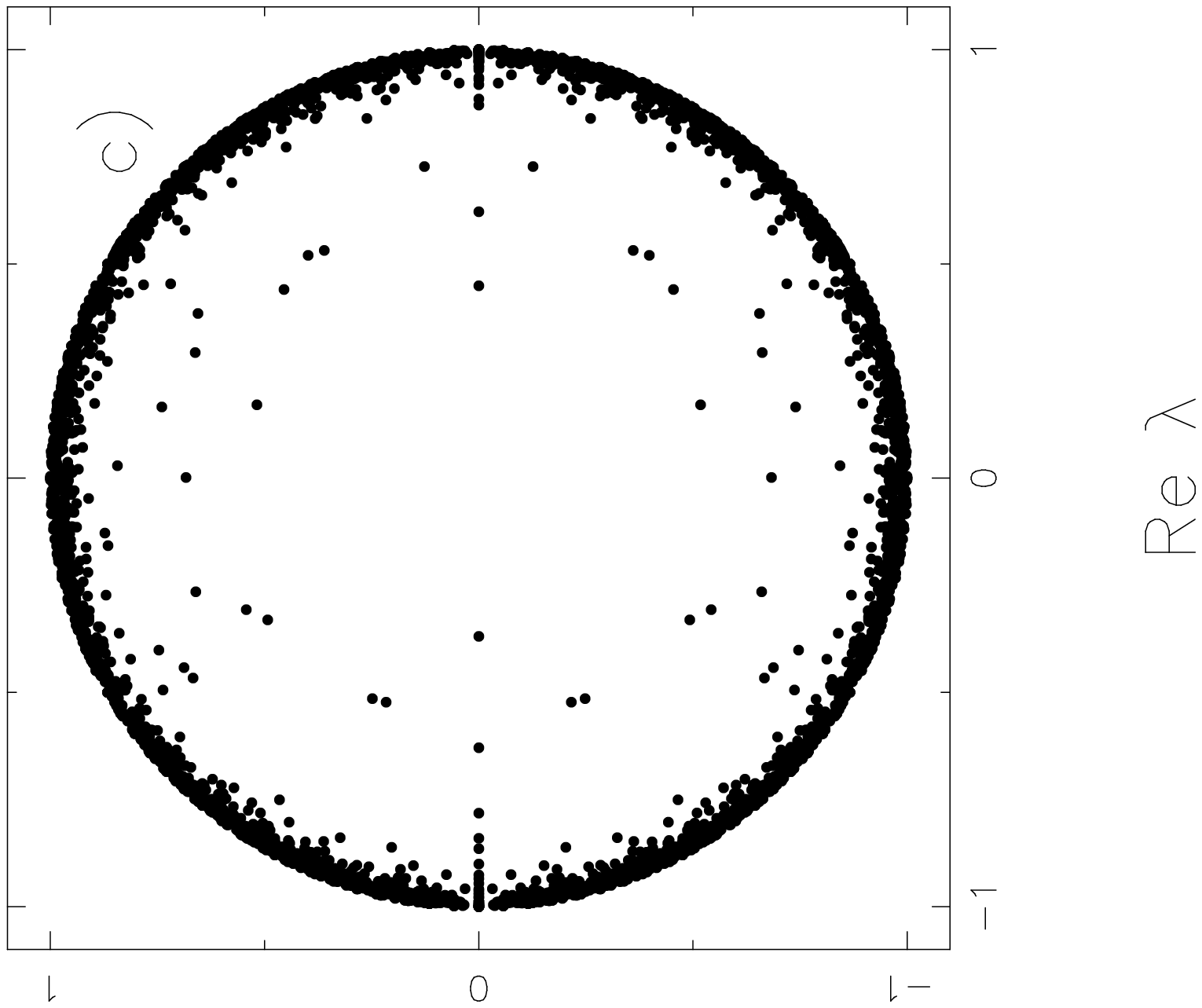}
\vspace{0.2cm}
\includegraphics*[width=5cm,angle=-90]{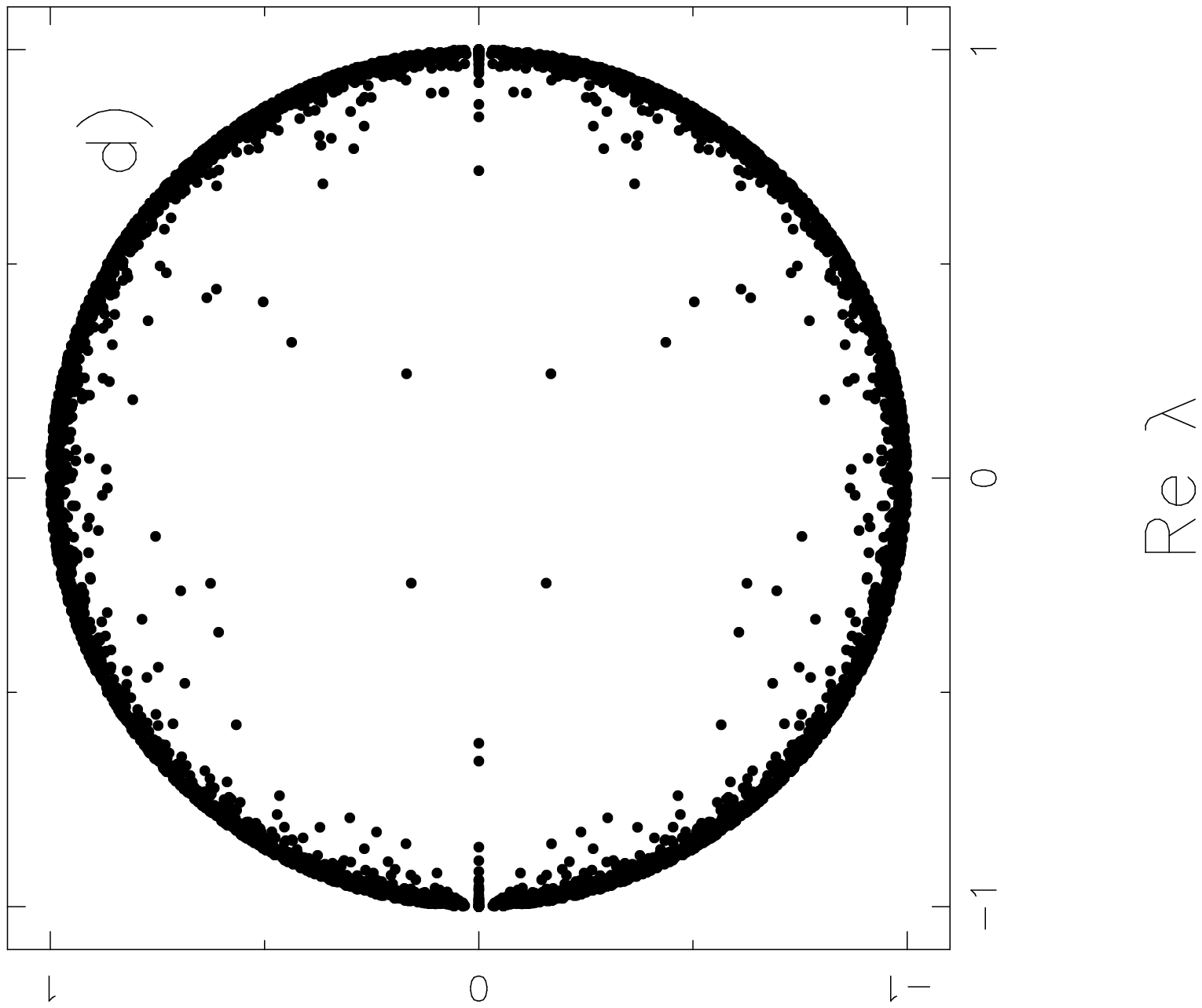}
\vspace{0.2cm}
\caption{%
Classical spectrum (Case II) for $l_{max}=20,40,60,70 $.
\label{spect2} }
\end{center}
\end{figure}

\pagebreak

\begin{figure}[h]
\begin{center}
\includegraphics*[width=8cm,angle=-90]{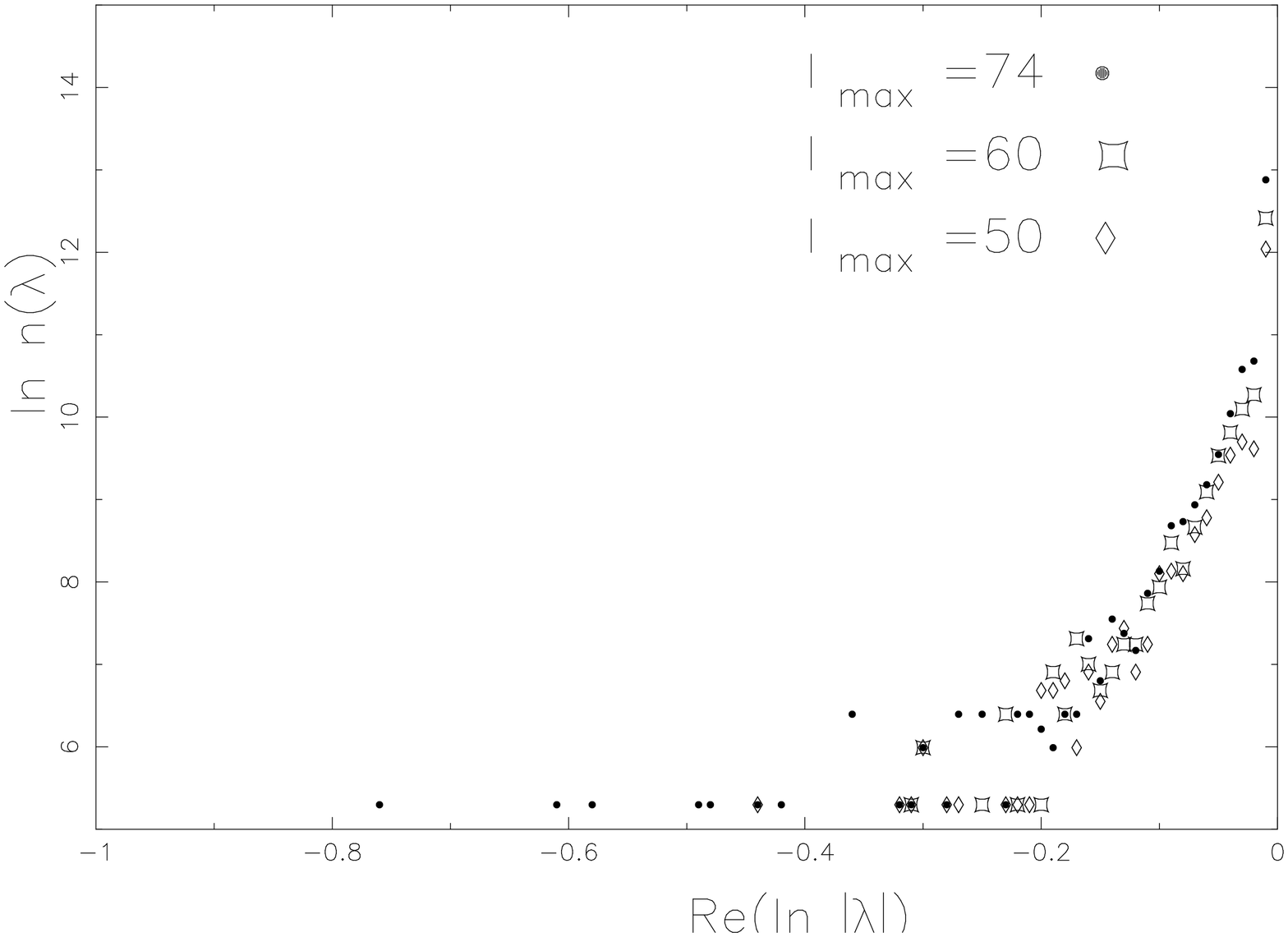}
\vspace{0.2cm}
\end{center}
\end{figure}

\pagebreak

\pagebreak

\end{document}